\documentclass[twocolumn,amsmath,footinbib,amssymb,aps,superscriptaddress]{revtex4-1}
\usepackage[latin9]{inputenc}
\usepackage{mathtools}
\usepackage{amsmath}
 
\usepackage[colorlinks=true,linktocpage,bookmarks=true,citecolor=blue,linkcolor=blue,urlcolor=blue,hyperfootnotes=true,pageanchor=true]{hyperref}
\usepackage[caption=false,labelformat=simple]{subfig} 
\usepackage{adjustbox}
\usepackage{multirow}

\makeatletter

\providecommand{\tabularnewline}{\\}

\usepackage{mathrsfs}
\usepackage{wasysym}\usepackage[mathscr]{eucal}
\usepackage{xfrac}
\usepackage{dcolumn}
\usepackage{bm}
\usepackage{color}
\usepackage{tabularx}
\usepackage[normalem]{ulem}
\usepackage[all]{hypcap}
\usepackage[dvipsnames,usenames,table]{xcolor}

\newcommand{\nocontentsline}[3]{}
\newcommand{\tocless}[2]{\bgroup\let\addcontentsline=\nocontentsline#1{#2}\egroup}

\newcommand{\be}{\begin{equation}}
\newcommand{\ee}{\end{equation}}
\newcommand{\bea}{\begin{eqnarray}}
\newcommand{\eea}{\end{eqnarray}}

\newcommand{\mc}{\mathcal}
\newcommand{\mb}{\mathbf}

\makeatother

\begin{document}
\title{Topological transition from nodal to nodeless Zeeman splitting in altermagnets}
\author{Rafael M. Fernandes}
\email{rfernand@umn.edu}
\affiliation{School of Physics and Astronomy, University of Minnesota, Minneapolis,
Minnesota 55455, USA}
\author{Vanuildo S. de Carvalho}
\affiliation{Instituto de F\'{\i}sica, Universidade Federal de Goi\'as, 74.001-970, Goi\^ania-GO,
Brazil}
\author{Turan Birol}
\affiliation{Department of Chemical Engineering and Materials Science, University
of Minnesota, Minneapolis, MN 55455, USA}
\author{Rodrigo G. Pereira}
\affiliation{International Institute of Physics and Departamento de F\'{\i}sica Te\'orica e Experimental, Universidade Federal
do Rio Grande do Norte, 59072-970 Natal-RN, Brazil}
\begin{abstract}

In an altermagnet, the symmetry that relates configurations with flipped magnetic moments is a rotation. This makes it qualitatively different from a ferromagnet, where no such symmetry exists, or a collinear antiferromagnet, where this symmetry is a lattice translation.
In this paper, we investigate the impact of the crystalline environment, enabled by the spin-orbit coupling, on the magnetic and electronic properties of an altermagnet. We find that, because each component of the magnetization acquires its own angular dependence, the Zeeman splitting of the bands has symmetry-protected nodal lines residing on mirror planes of the crystal. Upon crossing the Fermi surface, these nodal lines give rise to pinch points that behave as single or double type-II Weyl nodes. We show that an external magnetic field perpendicular to these mirror planes can only move the nodal lines, such that a critical field value is necessary to collapse
the nodes and make the Weyl pinch points annihilate. This unveils
the topological nature of the transition from a nodal to a nodeless Zeeman splitting of the bands. We also classify the
altermagnetic states of common crystallographic point groups in the presence of spin-orbit coupling,
revealing that a broad family of magnetic orthorhombic perovskites
can realize altermagnetism. 
\end{abstract}
\date{\today}

\maketitle

\section{Introduction}

Altermagnetism refers to a broad range of magnetically ordered states
that cannot be described in terms of standard ferromagnetic (F) or
antiferromagnetic (AF) orders \citep{Smejkal2020,Smejkal2022_1,Smejkal2022_2,Turek2022,Mazin2021,Feng2022,Betancourt2023,Mazin2023,Bai2023,Aoyama2023,Facio2023,Autieri2023,Okamoto2023,Steward2023}.
The distinguishing feature between these states is the type of crystalline
symmetries that leave their spin configuration unchanged when combined
with time reversal, i.e., with flipping the spins \cite{Smejkal2022_1,Smejkal2022_2}. In a ferromagnet,
there is no such symmetry, hence the material acquires a non-zero
magnetization. In a collinear antiferromagnet, translation by a lattice
vector ``undoes'' time reversal, resulting in a non-zero staggered
magnetization. In contrast, a lattice translation or inversion alone cannot undo the flipping of the spins of an altermagnetic
(AM) state, but a rotation with respect to
an axis or reflection with respect to a plane can. These distinct symmetry
properties lead to important consequences, most notably on the Zeeman
splitting of the band structure. To illustrate the differences between F, AF, and AM states, consider the following parametrization of the local magnetization $\mathbf{M}(\mathbf{r})$
\begin{equation}
\mathbf{M}\left(\mathbf{r}\right)=\mathbf{M}_{0}\,d\left(\hat{\mathbf{r}}\right)\cos\left(\mathbf{Q}\cdot\mathbf{r}\right)\ ,\label{eq:M_noSOC}
\end{equation}
where $d\left(\hat{\mathbf{r}}\right)$ is a function that depends
only on the direction $\hat{\mathbf{r}}$ and $\mathbf{Q}$ is the
magnetic wave-vector. Upon time-reversal, $\mathbf{M}_{0}\rightarrow-\mathbf{M}_{0}$
by definition. When $\mathbf{Q}=\mathbf{0}$
and $d\left(\hat{\mathbf{r}}\right)=1$, it is not possible to undo
the sign change of $\mathbf{M}_{0}$, and one has a ferromagnet. In this case, the spin degeneracy of
the electronic states is lifted, resulting in a Zeeman splitting between the
spin-up and spin-down bands. On the other hand, in the case of a commensurate
collinear antiferromagnet for which $d\left(\hat{\mathbf{r}}\right)=1$
and $2\mathbf{Q}$ is a reciprocal lattice vector, the sign change
of $\mathbf{M}_{0}$ imposed by time reversal can be compensated by
a translation by the appropriate lattice vector $\mathbf{R}$, since
$\cos\left(\mathbf{Q}\cdot\mathbf{R}\right)=-1$. Therefore, the spin
degeneracy is preserved, and the bands are not subjected to Zeeman
splitting in the AF state. 

There are other magnetic configurations given by Eq. (\ref{eq:M_noSOC}), however, that cannot be described as either a F or an AF state. Consider, for example, the case where the wave-vector is trivial, $\mathbf{Q}=\mathbf{0}$,
but the form factor $d\left(\hat{\mathbf{r}}\right)$ is not, i.e.
there is a crystalline symmetry operation $\mathcal{R}$ such that
$d\left(\mathcal{R}\hat{\mathbf{r}}\right)=-d\left(\hat{\mathbf{r}}\right)$.
For a centrosymmetric crystal, as long as this operation is not inversion (e.g. a rotation or a reflection),
the spin degeneracy is lifted, causing a Zeeman splitting in the
band structure that is neither uniform nor requires the presence of spin-orbit coupling (SOC). The resulting state is an example of an altermagnet \citep{Smejkal2022_1,Smejkal2022_2,Turek2022,Facio2023}, a classification that also encompasses a range of magnetic systems that display non-relativistic
Zeeman spin-splitting \citep{Kusunose2019,Kunes2019,Zunger2020,Zunger2022,Spaldin2022}. This distinction between F, AF, and AM states can be cast in formal grounds in terms of the three distinct types of spin groups -- generalizations of magnetic groups in which rotations in spin space are decoupled from real space operations \cite{Smejkal2022_1}.

To shed further light on the nature of altermagnetism, and on its
connection with other concepts of many-body electronic systems \citep{Smejkal2022_2},
it is useful to consider the special case of an isotropic system,
for which $d\left(\hat{\mathbf{r}}\right)$ can be expressed in terms
of spherical harmonics $Y_{lm}\left(\hat{\mathbf{r}}\right)$. In the situation where $l$ is
positive and even, the AM order parameter
in Eq. (\ref{eq:M_noSOC}) can be understood as a magnetization with
a non-zero angular momentum, corresponding to e.g. a $d$-wave or
a $g$-wave ``ferromagnet'' \citep{Kunes2019}. These types of order parameter, in turn,
naturally emerge within the well-understood Pomeranchuk instabilities of a
Fermi liquid in the spin-triplet $l=2$ or $l=4$ channels, respectively
\citep{Pomeranchuk1958}. Therefore, any even-parity, spin-triplet $l>0$ Pomeranchuk instability of a metal, whose general properties were investigated in Ref. \citep{Fradkin2007}, results in an altermagnet; note, however, that an AM state can also be realized in insulators. An appealing realization of such a spin-triplet Pomeranchuk instability is the so-called nematic-spin-nematic state \citep{Oganesyan2001,Fradkin2007}, which corresponds to a $d$-wave modulation of the spin polarization,
and displays unique collective modes.

The parametrization of $d\left(\hat{\mathbf{r}}\right)$
in terms of spherical harmonics also allows us to conclude, via a straightforward
calculation (see Appendix \ref{sec_Multipolar}), that the AM magnetic configuration described by Eq. (\ref{eq:M_noSOC})
displays a higher-order magnetic multipole moment of $l+1$ rank --
i.e., a magnetic octupole for $l=2$ or a magnetic dotriacontapole
for $l=4$, as opposed to the magnetic dipole condensed in ferromagnets
($l=0$) \citep{Kusunose2018,Spaldin2022}. Multipolar orders \citep{Winkler2023,Mosca2022,Spaldin2007,Hayami2018}
have been a common theme in studies of correlated $f$-electron systems
\citep{Kusunose2008,Santini2009} and $d$-electron systems \citep{Fu2015,Paramekanti2020}
with strongly-coupled spin and orbital degrees of freedom. In the case of AM, the multipole moments can be understood as arising from the multipole expansion of the electronic spin density rather than from the electronic configuration of an isolated atom.

Finally,
within a more microscopic description, the coarse-grained function
$d\left(\hat{\mathbf{r}}\right)$ can be attributed to intra-unit-cell
``antiferromagnetism'', i.e. a non-trivial configuration of the
magnetic moments of the atoms in a unit cell that yields a zero net
magnetization and that does not break translational symmetry \citep{Kusunose2019,Zunger2022}. It is important that the symmetry connecting opposite magnetic moments within the unit cell is not inversion, in which case one obtains instead of an altermagnet a compensated antiferromagnet with spin-degenerate bands (see, for example, the case of CuMnAs in Ref. \cite{Smejkal2017}). The latter, in turn, is described by the same type of spin groups as an AF with non-zero wave-vector \cite{Smejkal2022_1}.

Thus, altermagnets have a close relationship with non-$s$-wave
``ferromagnets,'' multipolar magnets, or intra-unit-cell ``antiferromagnets.''
Previous works, which provided crucial insight about the properties
of altermagnets, have primarily focused on the case in which the vector
$\mathbf{M}_{0}$ in Eq. (\ref{eq:M_noSOC}) can point in any direction
\citep{Smejkal2022_1,Smejkal2022_2,Turek2022,Liu2022,Yang2021}. Meanwhile, because of the ubiquitous presence of the SOC, the crystalline environment inevitably restricts the possible directions of the magnetic moments, even when the SOC is weak. Here, we investigate the impact of the coupling to the crystalline environment, and thus of the SOC, on the properties of an
altermagnet. Our main finding is that the
local magnetization in an altermagnet is generally not collinear,
and that Eq. (\ref{eq:M_noSOC}) must be replaced by
\begin{equation}
\mathbf{M}\left(\mathbf{r}\right)=M_{0}\,\mathbf{d}\left(\hat{\mathbf{r}}\right)\cos\left(\mathbf{Q}\cdot\mathbf{r}\right)\ .\label{eq:M_SOC}
\end{equation}
The key point is that all three magnetization
components acquire their own angular dependences rather than the same
angular dependence as in Eq. (\ref{eq:M_noSOC}). We use group theory to determine the properties of $\mathbf{d}\left(\hat{\mathbf{r}}\right)$
for common crystallographic point groups, and illustrate it for the
candidate AM compound MnF$_{2}$. We also use these results to show
that one of the magnetic phases proposed to be stabilized in orthorhombic
perovskites with space group $Pnma$ is actually an AM phase that is not accompanied by a finite magnetization, thus
opening a new avenue to search for altermagnets that do not display an anomalous Hall effect. 
We emphasize that Eq. (\ref{eq:M_noSOC}) is a special case of Eq.
(\ref{eq:M_SOC}), and that the two formulations do not contradict
each other. In fact, as we show below, along high-symmetry crystallographic
planes, only one of the components of $\mathbf{d}\left(\hat{\mathbf{r}}\right)$
is non-zero, resulting in an effective collinear AM configuration (see also Ref. \citep{Facio2023}).

To assess the impact of these results on the band structure of altermagnets,
we solve the appropriate low-energy Hamiltonian and show that nodal
lines emerge, along which the Zeeman splitting between the bands vanish.
Remarkably, although topologically trivial with respect to non-spatial
symmetries, these nodal lines lie on mirror planes of the crystal,
and are thus protected by crystalline symmetries \citep{Schnyder2014}.
In altermagnetic metals, the Fermi surface acquires a non-trivial
spin texture and is split in two everywhere except at the pinch points
originating from the intersection with the nodal lines. Upon expanding
the Hamiltonian around these pinch points and calculating their Berry
phase, we find that the pinch points actually behave as type-II single or double Weyl nodes \citep{Bernevig2015,Bernevig2012}.

The symmetry-protection of the nodal lines has crucial implications
for the nature of the transition from an AM phase to a F phase driven
by a uniform magnetic field. At first sight, one might have expected
that the magnetic field would immediately generate a Zeeman splitting
everywhere in the band structure, thus destroying the Fermi-surface
pinch points that characterize AM metals. However, the nodal lines
cannot be destroyed by an infinitesimal field that is perpendicular
to the mirror plane along which the nodal lines lie. Instead, a small
magnetic field leads to closed nodal loops that move along the mirror
plane. Consequently, a fully Zeeman-split band structure, which is
characteristic of a ferromagnetic phase, only emerges for a large
enough magnetic field, which is necessary to collapse the nodal loops.
Conversely, the Weyl-like pinch points of the Fermi surface located on
the plane perpendicular to the field move towards each other and annihilate
for a critical value of the field. Thus, the AM-F transition in which
the Zeeman splitting of the bands changes from nodal to nodeless is
a topological transition.

\begin{table*}
\begin{centering}
\begin{tabular}{|c|c|c|}
\hline
\hline 
Point group & AM irrep & $\mathbf{d}_{i}\left(\mathbf{k}\right)\equiv\left(d_{i,x},\,d_{i,y},\,d_{i,z}\right)$\tabularnewline
\hline  
$D_{2h}$ & $A_{1g}^{-}$ & $\left(k_{y}k_{z},\,\eta_{1}k_{x}k_{z},\,\eta_{2}k_{x}k_{y}\right)$\tabularnewline
\hline 
\multirow{3}{*}{$D_{4h}$} & $A_{1g}^{-}$ & $\left(k_{y}k_{z},-k_{x}k_{z},\,\eta\,k_{x}k_{y}\left(k_{x}^{2}-k_{y}^{2}\right)\right)$\tabularnewline
\cline{2-3} 
        & $B_{1g}^{-}$ & $\left(k_{y}k_{z},\,k_{x}k_{z},\,\eta\,k_{x}k_{y}\right)$\tabularnewline
\cline{2-3} 
        & $B_{2g}^{-}$ & $\left(-k_{x}k_{z},\,k_{y}k_{z},\,\eta\left(k_{x}^{2}-k_{y}^{2}\right)\right)$\tabularnewline
\hline 
\multirow{4}{*}{$D_{6h}$} & $A_{1g}^{-}$ & $\left(k_{y}k_{z},-k_{x}k_{z},\,\eta\,k_{x}k_{y}\left(k_{x}^{2}-3k_{y}^{2}\right)\left(3k_{x}^{2}-k_{y}^{2}\right)\right)$\tabularnewline
\cline{2-3} 
        & $B_{1g}^{-}$ & $\left(k_{x}^{2}-k_{y}^{2},-2k_{x}k_{y},\,\eta\,k_{x}k_{z}\left(k_{x}^{2}-3k_{y}^{2}\right)\right)$\tabularnewline
\cline{2-3} 
        & $B_{2g}^{-}$ & $\left(2k_{x}k_{y},\,k_{x}^{2}-k_{y}^{2},\,\eta\,k_{y}k_{z}\left(3k_{x}^{2}-k_{y}^{2}\right)\right)$\tabularnewline
\cline{2-3} 

        & $E_{2g}^{-}$ & $\begin{cases}
\left(k_{y}k_{z},\,k_{x}k_{z},\,2\eta\,k_{x}k_{y}\right) & \:,\,i=1\\
\left(k_{x}k_{z},-k_{y}k_{z},\eta\left(k_{x}^{2}-k_{y}^{2}\right)\right) & \:,\,i=2
\end{cases}$\tabularnewline

\hline 
\multirow{4}{*}{$O_{h}$} & $A_{1g}^{-}$ & $\left(k_{y}k_{z}\left(k_{y}^{2}-k_{z}^{2}\right),\,k_{x}k_{z}\left(k_{z}^{2}-k_{x}^{2}\right),\,k_{x}k_{y}\left(k_{x}^{2}-k_{y}^{2}\right)\right)$\tabularnewline
\cline{2-3} 
        & $A_{2g}^{-}$ & $\left(k_{y}k_{z},\,k_{x}k_{z},\,k_{x}k_{y}\right)$\tabularnewline
\cline{2-3} 
        & $E_{g}^{-}$ & $\begin{cases}
\sqrt{3}\left(k_{y}k_{z},\,-k_{x}k_{z},\,0\right) & \:,\,i=1\\
\left(-k_{y}k_{z},-k_{x}k_{z},\,2k_{x}k_{y}\right) & \:,\,i=2
\end{cases}$\tabularnewline
\cline{2-3} 
        & $T_{2g}^{-}$ & $\begin{cases}
\left(k_{x}k_{z},-k_{y}k_{z},\,\eta \left( k_x^2 - k_y^2 \right) \right) & \:,\,i=1\\
\left(\eta \left( k_y^2 - k_z^2 \right),\,k_{x}k_{y},-k_{x}k_{z}\right) & \:,\,i=2\\
\left(-k_{x}k_{y},\,\eta \left( k_z^2 - k_x^2 \right),k_{y}k_{z}\right) & \:,\,i=3
\end{cases}$\tabularnewline
\hline 
\hline
\end{tabular}
\par\end{centering}
\caption{Classification of the \textquotedblleft pure\textquotedblright{} altermagnetic (AM)
order parameters in the four point groups $D_{2h}$ (orthorhombic),
$D_{4h}$ (tetragonal), $D_{6h}$ (hexagonal), and $O_{h}$ (cubic).
The minus sign at the superscript of the irreducible representations
(irreps) indicates that the order parameter is odd under time reversal.
The third column shows the small-momentum expansion of the vector
$\mathbf{d}_{i}\left(\mathbf{k}\right)$ associated with each of the
allowed AM order parameters. The parameters $\eta_{i}$ are related
to magnetic anisotropy and cannot be determined based on symmetry
alone. \label{tab:classification}}
\end{table*}

This paper is organized as follows: in Sec. \ref{sec:AM_SOC}, we present a group-theoretical classification of AM states in the presence of SOC. Sec. \ref{sec:model} introduces the effective low-energy models for AM and demonstrates their intrinsic non-collinearity. In Sec. \ref{sec:nodal_lines}, we show the emergence of symmetry-protected nodal lines in the Zeeman splitting of AM. The topological character of the transition from an altermagnetic to a ferromagnetic state is discussed in \ref{sec:topo_transition}. Sec. \ref{sec:concl} is devoted to the conclusions, whereas details of the model and of the calculations discussed in the main text are presented in Appendices \ref{sec_Multipolar}-\ref{sec_NodalLines}. 

\section{Altermagnets in the presence of SOC}\label{sec:AM_SOC}

We start by employing group theory to classify the AM order parameter
$\Phi^{i}$ (where $i$ denotes different components) whose underlying crystalline environment, via the SOC, forces the magnetic moments to point along certain directions. For the magnetization $\boldsymbol{m}$, this implies that it must transform as the irreducible representations (irreps)
of the crystallographic point group, rather than a vector in the space
of rotations. For instance, in the tetragonal group $D_{4h}$, $m_{z}$
transforms as the one-dimensional (1D) irrep $A_{2g}^{-}$, whereas
$\left(m_{x},\,m_{y}\right)$ transform as the two-dimensional (2D)
irrep $E_{g}^{-}$. Here, the minus superscript indicates that the
quantity is odd under time reversal. The situation is analogous in
the case of $\Phi^{i}$. In the widely studied cases of AM states that preserve inversion and have zero wave-vector, $\Phi^{i}$ must
transform as one of the time-reversal-odd, even-parity irreps of a centrosymmetric
point group. We also add the constraint that the AM phase does not
display a non-zero magnetization, like a ferromagnet does. This excludes
the irreps that transform as dipolar magnetic moments, such as the
aforementioned $A_{2g}^{-}$ and $E_{g}^{-}$ in the tetragonal group
$D_{4h}$. These ``pure'' AM phases should be distinguished from
mixed AM-F phases, in which the onset of $\Phi^{i}$ necessarily triggers
a non-zero magnetic moment. We will return to this point later.

The irreps of the AM order parameter for common point groups -- orthorhombic
$D_{2h}$, tetragonal $D_{4h}$, hexagonal $D_{6h}$, and cubic $O_{h}$
-- are shown in the first two columns of Table \ref{tab:classification}.
Extension to other point groups is straightforward. In most cases,
$\Phi$ transforms as a 1D irrep, and thus behaves as an Ising-like
order parameter \cite{Steward2023}. The exceptions are the 2D irreps $E_{2g}^{-}$ of
the hexagonal group and $E_{g}^{-}$ of the cubic group, where $\left(\Phi^{1},\Phi^{2}\right)$
behaves as a six-state clock-model order parameter, as well as the
3D irrep $T_{2g}^{-}$ of the cubic group, in which $\left(\Phi^{1},\Phi^{2},\Phi^{3}\right)$
has the same free-energy as that of a Heisenberg ferromagnet with
a cubic anisotropy term (see Appendix \ref{sec_Landau}). This analysis also makes the aforementioned
connection between AM order and magnetic multipolar order explicit.
Indeed, by using the crystallographic classification of multipoles
of Ref. \citep{Kusunose2018}, one readily identifies the correspondence
between $\Phi^{i}$ and different types of multipole moments of the electronic wave-function. For
instance, $\Phi$ transforming as $B_{1g}^{-}$ in the tetragonal
group is equivalent to a magnetic octupole moment, as pointed out
in Ref. \citep{Spaldin2022}, whereas $\left(\Phi^{1},\Phi^{2},\Phi^{3}\right)$
transforming as $T_{2g}^{-}$ in the cubic group corresponds to magnetic
quadrupolar toroidal moments.

\begin{figure}
    \centering
    \includegraphics[width=0.60\linewidth]{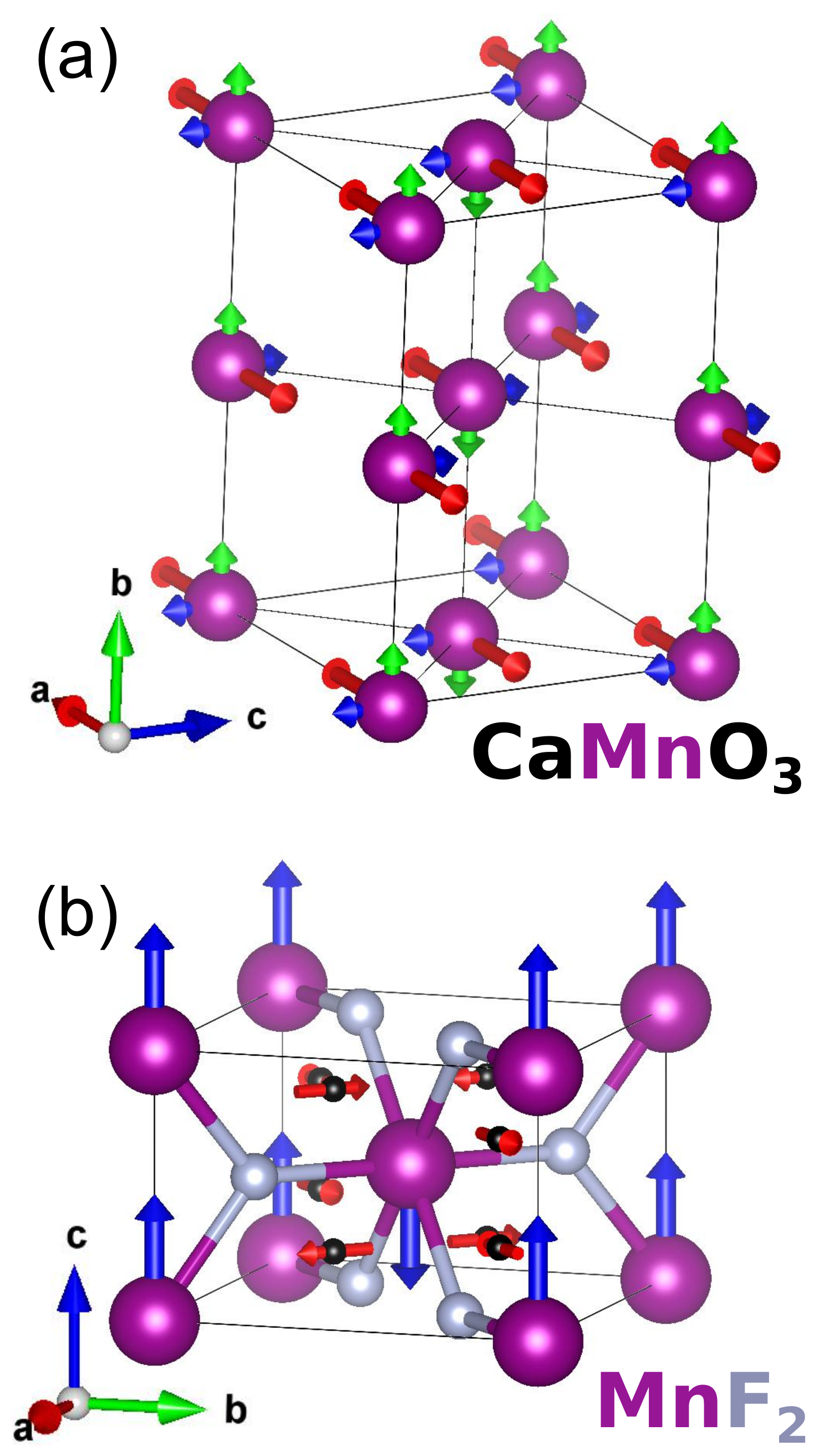}
    \caption{\label{fig:Materials} \textbf{Non-collinearity of the magnetic moments in altermagnetic phases}. (a) The primitive unit cell of $Pnma$ CaMnO$_3$ contains four Mn ions (purple spheres). If G-type intra-unit-cell AF order along the $a$ axis is present (the $G_a$ order, shown as red arrows), the B-site moments are allowed to be tilted without reducing the symmetry. This leads to a C-type intra-unit-cell AF order along the $b$ axis (the $C_b$ order, green arrows), and an A-type order along the $c$ axis (the $A_c$ order, blue arrows). The resulting $G_aC_bA_c$ order is an altermagnetic state. The relative amplitude between the arrows were obtained from first principles calculations (see Appendices \ref{sec_Abinitio_A}-\ref{sec_Abinitio_C}).
    (b) The unit cell of MnF$_2$ contains 2 Mn ions (purple) and 4 F ions (grey). In the altermagnetic phase, the magnetic moments of the corner and the body-center Mn moments are antiparallel along the $c$ axis. While all atomic moments are along $c$, the spin-density on the 8j Wyckoff site (shown as small black spheres) is non-collinear and displays the in-plane components shown in the figure. }
\end{figure}

Besides providing a sound framework to investigate the properties
of altermagnets, which we pursue below, this classification also offers
interesting insights about candidate AM materials. 
Table \ref{tab:classification} shows that, for orthorhombic
crystals, there is only one allowed even-parity ``pure'' AM order parameter $\Phi$, which
transforms as $A_{1g}^{-}$. Interestingly, there is a broad class
of orthorhombic materials known to display $\mathbf{Q}=0$ magnetism:
ABO$_{3}$ perovskites with space group $Pnma$, such as rare-earth
titanates (LaTiO$_3$) and manganites (CaMnO$_3$) \cite{Schmitz2005,Spaldin2011}. In these systems, the ideal perovskite cubic
crystal structure is distorted into an orthorhombic one by octahedral
rotations that emerge to accommodate the A-site atom. Because of the
expanded unit cell of the orthorhombic crystal structure shown in Fig.~\ref{fig:Materials}(a), magnetic
orders characterized by a finite wave-vector in the ideal cubic structure
become $\mathbf{Q}=0$ magnetic orders in the $Pnma$ space group.
The transformation properties of these intra-unit-cell ``antiferromagnetic''
states are well understood \citep{Schmitz2005,Spaldin2011,Zhentao2022}.
Comparing them with our classification of AM orders, we identify the
so-called $G_aC_bA_c$ magnetic state \citep{Zhentao2022} shown in Fig. \ref{fig:Materials}(a) (or $A_{a}G_{b}C_{c}$ if $Pbnm$ axes are used for the orthorhombic cell) as
an altermagnetic phase. In this notation, $G_a$ means there is a $G$-type ordering (wave-vector $(\pi,\pi,\pi)$ in the pseudocubic perovskite Brillouin zone) with magnetic moments pointing along the orthorhombic $a$-axis, superimposed with a $C$-type ordering (wave-vector $(\pi,\pi,0)$) with magnetic moments along the $b$-axis ($C_b$) and an $A$-type ordering (wave-vector $(0,0,\pi)$) with magnetic moments along the $c$-axis ($A_c$). While the known rare-earth
titanates realize a different magnetic state with a non-zero ferromagnetic moment, dubbed $A_aF_bG_c$ in the notation above, DFT calculations show that
the ground state energy of the AM $G_aC_bA_c$ phase is a close competitor \citep{Schmitz2005}. Note that, as it follows from the analyses of Refs. \cite{Smejkal2020,Samanta2020}, the ground state $A_aF_bG_c$ with a dominant $G_c$ component corresponds to an AM phase with a weak-ferromagnetic component \cite{Bousquet2011, Birol2012}, which we here dubbed a mixed AM-F phase. The main difference with respect to the $G_aC_bA_c$ state is that only the former displays an anomalous Hall effect, which is observed experimentally in doped CaMnO$_3$ \cite{Vistoli2019} and is related to the fact that the symmetries of that AM configuration allow for a non-zero macroscopic magnetization \cite{Makoto2022}.

\section{Low-energy models for altermagnets}\label{sec:model}

While our group-theoretical results are not restricted to a particular
model or material, in order to elucidate the coupling between $\Phi^{i}$
and the electronic degrees of freedom, it is useful to consider a
general low-energy model. Thus, we start from a non-interacting single-band
Hamiltonian $\mathcal{H}_{0}=\sum_{\mathbf{k}}\varepsilon_{\mathbf{k}}c_{\mathbf{k}s}^{\dagger}c_{\mathbf{k}s}^{\phantom{\dagger}}$,
with electronic dispersion $\varepsilon_{\mathbf{k}}$ and electronic
operator $c_{\mathbf{k}s}$, where $\mathbf{k}$ is the momentum and
$s$ is the spin (summation over the spin indices is implicitly assumed
in this paper). For simplicity, we assume that the orbital described
by $c_{\mathbf{k}s}$ transforms as a one-dimensional irreducible
representation (irrep) of the point group; generalizations to multi-orbital
models are straightforward. The AM order parameter $\Phi^{i}$ must
couple to a fermionic bilinear that transforms as the same irrep as
$\Phi^{i}$. To construct these bilinears, we extend the parametrization
of the spatial-dependent magnetization in Eq. (\ref{eq:M_noSOC})
to the momentum-dependent quantum spin-density, which is accomplished
upon replacing $\mathbf{M}_{0}$ by the operator $c_{\mathbf{k}s}^{\dagger}\boldsymbol{\sigma}_{ss'}c_{\mathbf{k}s'}^{\phantom{\dagger}}$,
where $\sigma_{ss'}^{\mu}$ are Pauli matrices (with $\mu=x,y,z$).
Then, the interacting Hamitonian $\mathcal{H}_{\mathrm{int}}$ that
describes the coupling between the AM order parameter and the electrons
is given by:

\begin{figure}
\centering
\centering \includegraphics[width=0.6\linewidth]{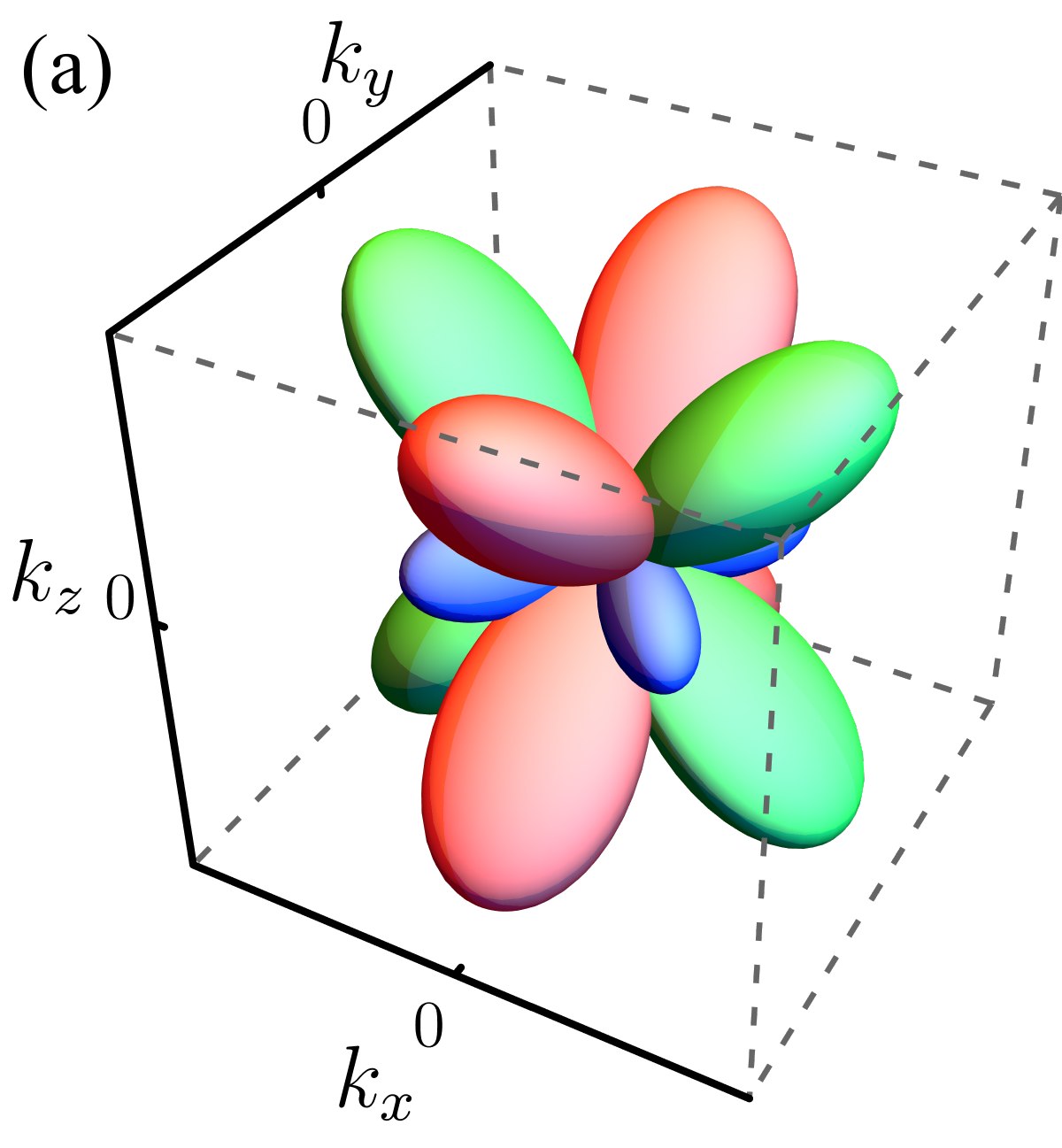} \vfil{}  \includegraphics[width=0.6\linewidth]{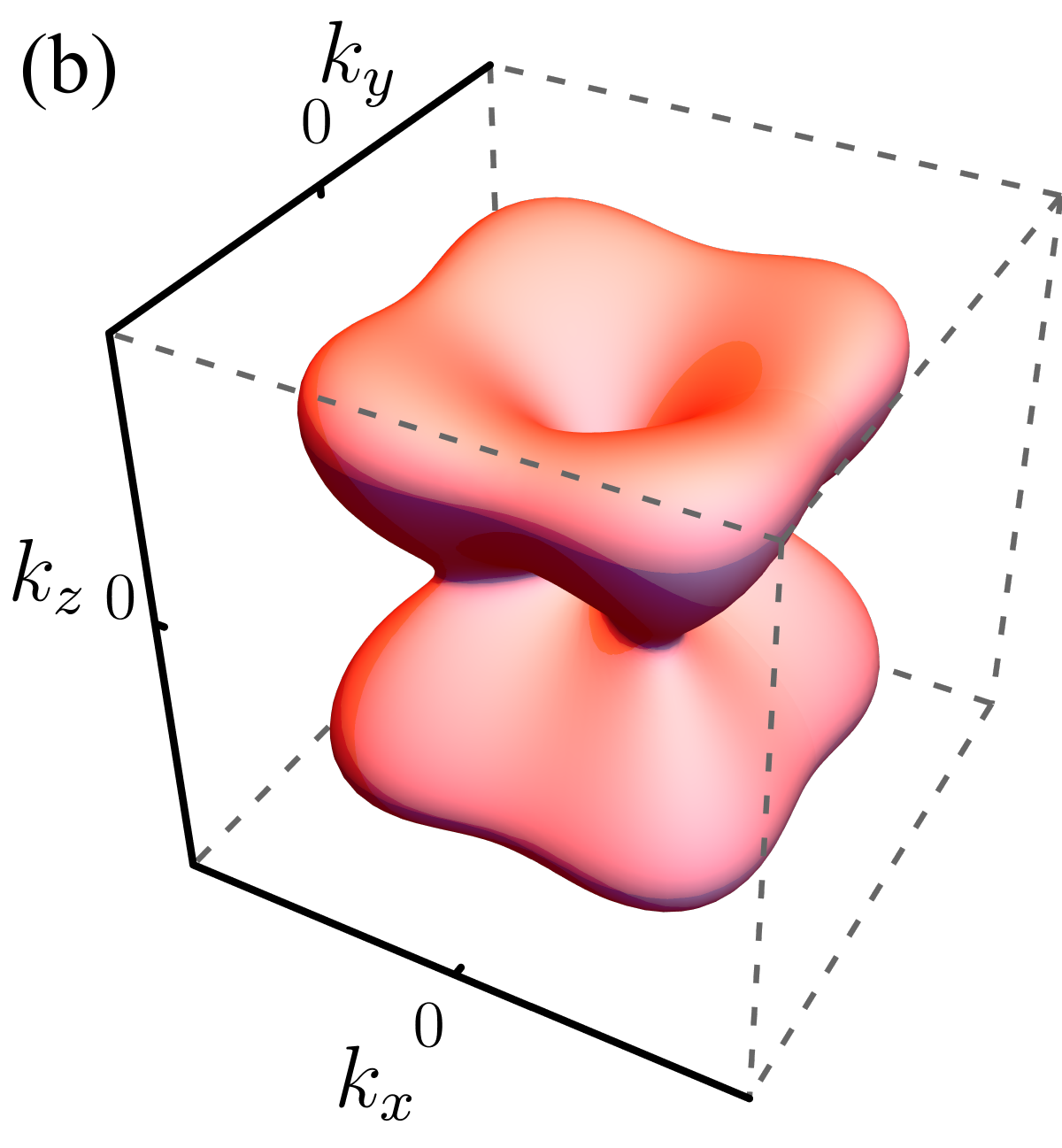} 
\caption{\textbf{Polar plot of the d-vector $\mathbf{d}\left(\mathbf{k}\right)$.} Panel (a) shows the squared three components $d_{\mu}^2$ (red corresponds to $d_{x}$, green to $d_y$, and blue to $d_z$) while panel (b) shows the total magnitude $|\mathbf{d}|^2$. Here, we considered the specific case of an AM order parameter that transforms as the $B_{1g}^{-}$ irrep of the tetragonal group $D_{4h}$, for which $\mathbf{d}\left(\mathbf{k}\right) = \left( k_y k_z,\, k_x k_z,\, \eta k_x k_y \right)$; we set $\eta = 3/4$ in this plot.}\label{fig:d_vectors}
\end{figure}

\begin{equation}
\mathcal{H}_{\mathrm{int}}=-\lambda\sum_{\mathbf{k},i}\Phi^{i}\left[\mathbf{d}_{i}\left(\mathbf{k}\right)\cdot\boldsymbol{\sigma}_{ss'}\right]c_{\mathbf{k}s}^{\dagger}c_{\mathbf{k}s'}^{\phantom{\dagger}}\ ,\label{eq:H_int}
\end{equation}
where $\lambda$ is a coupling constant and the dot product refers
to spin space, i.e. $\mathbf{d}_{i}\left(\mathbf{k}\right)\cdot\boldsymbol{\sigma}_{ss'}=\sum_{\mu}d_{i,\mu}\left(\mathbf{k}\right)\sigma_{ss'}^{\mu}$. 
This is the analogue of Eq. (\ref{eq:M_SOC}) in momentum space. Recall that the index $i$ refers to the number of components of the irrep that describes the AM order parameter. Since
the transformation properties of $\Phi^{i}$ and of $\boldsymbol{\sigma}$
in terms of the irreps of each group are known, it is straightforward
to determine the transformation properties of the set of vectors $\mathbf{d}_{i}\left(\mathbf{k}\right)$. More specifically, the procedure is as follows: for a given point group $\mathcal{G}$, we know the time-reversal-odd irreps $\Gamma_\sigma^-$ according to which the components of $\boldsymbol{\sigma}$ transform. Then, for a given AM order parameter $\Phi^{i}$ that transforms as one of the time-reversal-odd irreps $\Gamma_\Phi^-$ of $\mathcal{G}$ that are different from $\Gamma_\sigma^-$, we determine the irreps $\Gamma_d^+$ of the components of $\mathbf{d}_i$ such that $\mathbf{d}_{i}\left(\mathbf{k}\right)\cdot\boldsymbol{\sigma}$ transforms as $\Phi^{i}$, i.e. $\Gamma_d^+ \otimes \Gamma_\sigma^- = \Gamma_\Phi^-$. Once $\Gamma_d^+$ is determined, we construct the polynomials in Table \ref{tab:classification} using group theory.

We note that the low-energy model in Eq. (\ref{eq:H_int}) has similarities with that employed in Ref. \citep{Fradkin2007} to investigate the properties of spin-triplet Pomeranchuk instabilities. Indeed, as we discussed above, ordered states resulting from even-parity spin-triplet Fermi liquid instabilities are altermagnets. While Ref. \citep{Fradkin2007} considered the case of isotropic systems, here our focus is on the impact of the crystalline lattice. 
We also note that Eq. (\ref{eq:H_int}) is similar to that proposed in Ref. \citep{Kusunose2020} to model spin-split
bands in antiferromagnets without SOC, in which case functions analogous
to $d_{i\mu}\left(\mathbf{k}\right)$ described electric multipole moments arising from different types
of intra-unit-cell ``antiferromagnetism''. The main difference
is that in Ref. \citep{Kusunose2020} the spin
coordinates transformed independently of the lattice.

Using the procedure described above, we obtain the irreps for each
component of $\mathbf{d}_{i}\left(\mathbf{k}\right)$ associated with
a given AM order parameter $\Phi^{i}$, as shown in the third column
of Table \ref{tab:classification} \citep{Hatch2003INVARIANTS,StokesInvariants}. Here, we consider only the lowest-order polynomials in momentum that give non-zero $d_{i,\mu}\left(\mathbf{k}\right)$; these polynomials can also be expressed in terms of lattice harmonics
by applying standard methods (see e.g. \citep{Platt2013}). One important
feature of $\mathbf{d}_{i}\left(\mathbf{k}\right)$ is the presence of a prefactor $\eta$ in some
of its components. This parameter, which cannot be determined solely
on symmetry grounds, is directly related to the magnetic anisotropy
of the lattice. To illustrate this point, consider the case of the
AM order parameter $\Phi$ that transforms as the $B_{1g}^{-}$ irrep
of the tetragonal group, which has been invoked to describe the rutile AM
candidates MnF$_{2}$ and RuO$_{2}$ \cite{Smejkal2022_1,Spaldin2022}. Clearly, the combination $\sigma_{z}k_{x}k_{y}$
transforms as $B_{1g}^{-}$, since $\sigma_{z}$ transforms as $A_{2g}^{-}$
and $k_{x}k_{y}$, as $B_{2g}$. However, there is another combination
of Pauli matrices and d-wave form-factors that also transforms as $B_{1g}^{-}$.
Using the fact that $\left(\sigma_{x},\sigma_{y}\right)$ transforms
as $E_{g}^{-}$ while the doublet $\left(k_{y}k_{z},-k_{x}k_{z}\right)$
transforms as $E_{g}$, we find that $\left(k_{y}k_{z}\sigma_{x}+k_{x}k_{z}\sigma_{y}\right)$
also transforms as $B_{1g}^{-}$. As a result, $\Phi$ must couple
to both $\sigma_{z}k_{x}k_{y}$ and to $\left(k_{y}k_{z}\sigma_{x}+k_{x}k_{z}\sigma_{y}\right)$,
but with different coupling constants. When expressed in terms of
Eq. (\ref{eq:H_int}), we obtain $\mathbf{d}\left(\mathbf{k}\right)$
as in Table \ref{tab:classification}, with the parameter $\eta$
corresponding precisely to the ratio between the two coupling constants. To illustrate the properties of $\mathbf{d}\left(\mathbf{k}\right)$, we show in Fig. \ref{fig:d_vectors} a polar plot of the square of its three components $d_{\mu}^2$, as well as of its total magnitude $|\mathbf{d}|^2$, in the case of an AM order parameter that transforms $B_{1g}^{-}$ in the tetragonal group $D_{4h}$. Note that, in the absence of SOC, only one of the components $d_{\mu}$ is non-zero. 

We note that, in the presence of SOC, depending on the direction of the magnetic moments, altermagnetism may also trigger ferromagnetism. Consider, for instance, the AM state discussed above in the context of the rutiles, which is characterized by the d-wave form factor $k_{x}k_{y}$. Instead of an out-of-plane moment, let us instead assume that the moments point in the plane. The resulting two-component AM order parameter $\boldsymbol{\Phi} \sim \left( k_x k_y \sigma_y, \, k_x k_y \sigma_x\right)$ transforms as the $E_g^-$ irrep of $D_{4h}$. However, because the in-plane uniform magnetization $\boldsymbol{m} \sim \left( \sigma_x, \, \sigma_y \right)$ also transforms as $E_g^-$, the in-plane components of the $\mathbf{d}_i\left(\mathbf{k}\right)$ vectors necessarily acquire trivial components proportional to the uniform magnetization, i.e. $\mathbf{d}_{1,\parallel}\left(\mathbf{k}\right)=\left(m_x, \, k_{x}k_{y}\right)$ and $\mathbf{d}_{2,\parallel}\left(\mathbf{k}\right)=\left(k_{x}k_{y}, m_y\right)$. We dub this
type of order in which altermagnetism induces weak ferromagnetism a mixed AM-F phase. Although we will not discuss them further in the remainder of this paper, we note that AM configurations that allow an admixture with a weak ferromagnetic moment have been widely studied -- indeed, this is the case of the RuO$_2$ compound in the presence of an in-plane field that is large enough to make the magnetic moments switch from out-of-plane to in-plane \cite{Smejkal2020}. Of course, the distinction between pure and mixed AM phases is only meaningful in the presence of SOC, as they are described by the same spin group. Experimentally, a crucial difference between the pure AM and mixed AM-F phases is that only in the latter the system displays anomalous Hall effect or spontaneous Kerr rotation \cite{Smejkal2022_1}, as those effects are enabled by the same symmetries that allow for the admixture with a weak ferromagnetic moment. Indeed, as discussed elsewhere \cite{Tsymbal2020,MacDonald2020}, symmetry enforces the anomalous Hall effect to always be accompanied by a non-zero magnetization.

\begin{figure}
\centering
\centering \includegraphics[width=0.7\linewidth]{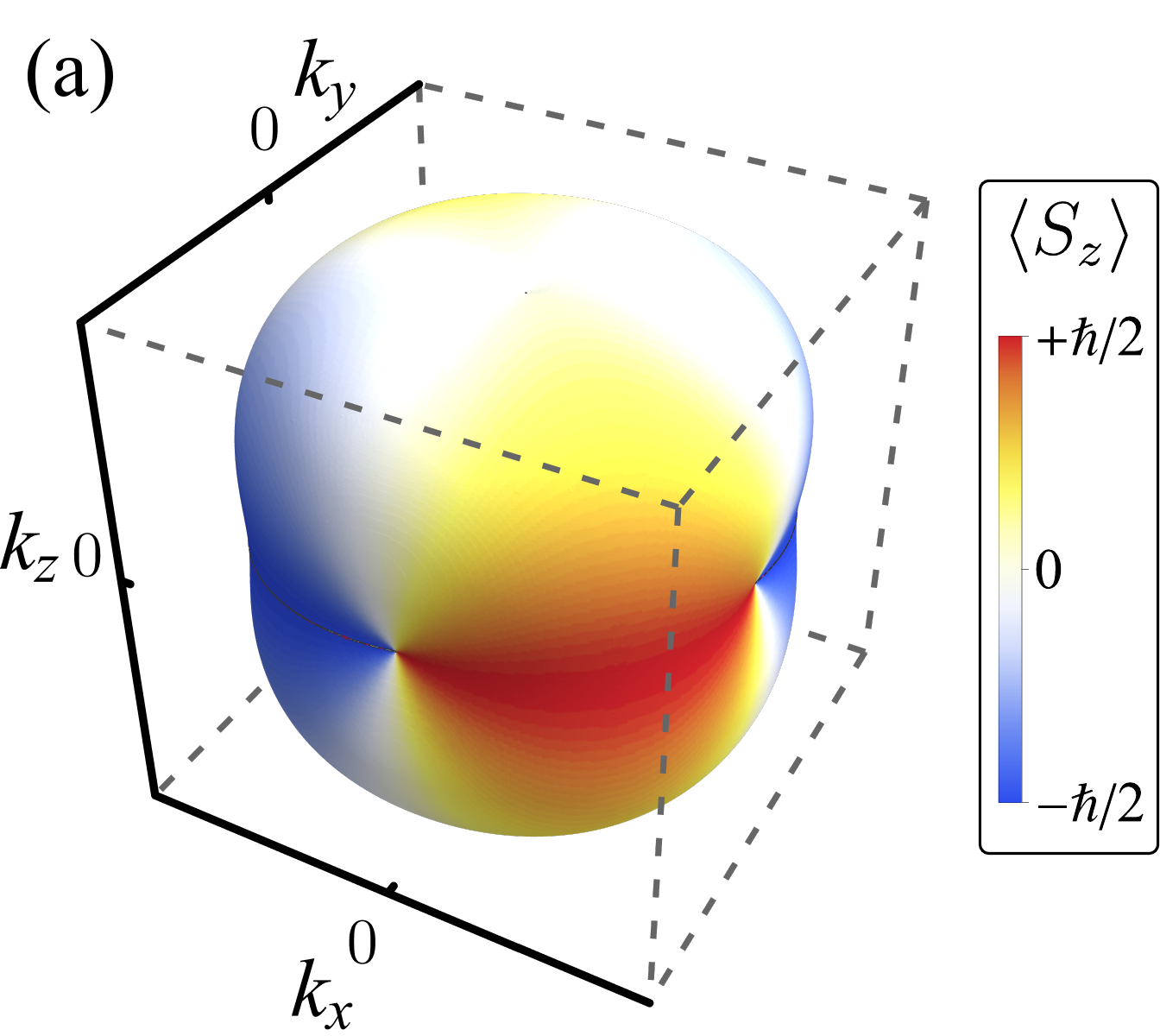} \vfil{}  \includegraphics[width=0.7\linewidth]{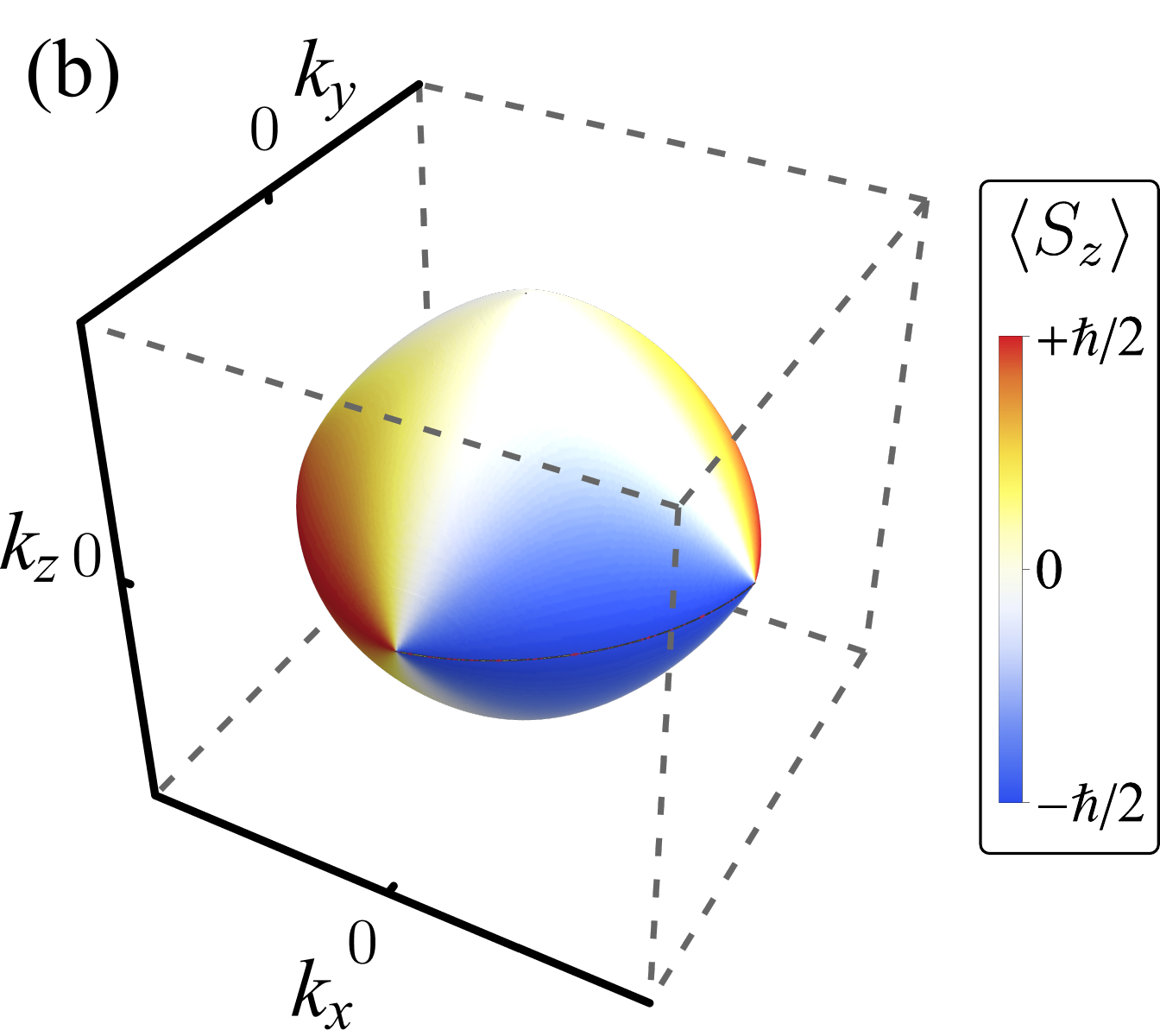} 
\caption{\textbf{Spin texture of the Fermi surface of an altermagnetic metal.} Expectation value of the out-of-plane component of the spin $S_z$ projected onto the two split Fermi surfaces of an altermagnetic metal, panels (a) and (b). Here, we consider the $B_{1g}^{-}$ AM state of a tetragonal system. The non-interacting Fermi surface is assumed spherical for simplicity.}\label{fig:spin_texture}
\end{figure}

The results displayed in Table \ref{tab:classification} show that
the vector $d_{i,\mu}\left(\mathbf{k}\right)$ never displays only
a single component $\mu$. Therefore, from Eq. (\ref{eq:H_int}),
we conclude that the AM order parameter corresponds to angular modulations
of more than one spin polarization -- or, in other words, that an
AM phase is not collinear when the magnetic and lattice degrees of freedom are coupled (see also Ref. \citep{Facio2023}). This points to the
fundamentally different role of SOC in AM and F systems, as in ferromagnets the magnetic anisotropy does not necessarily enforce non-collinearity.

In momentum space, the non-collinearity of the AM phase is manifested as a spin texture of the band structure. To illustrate this point, in Fig. \ref{fig:spin_texture} we show the expectation value of the out-of-plane spin component $S_z = \hbar \sigma_z/2$ projected onto the split Fermi surfaces inside the $B_{1g}^{-}$ AM state of a tetragonal system. This result is obtained by diagonalizing the Hamiltonian containing Eq. (\ref{eq:H_int}) for a non-zero $\Phi$; for simplicity, we consider a non-interacting spherical Fermi surface. Clearly, the split Fermi surfaces display a rich spin texture.
Nevertheless, we emphasize that along high-symmetry planes $\mathbf{d}\left(\mathbf{k}\right)$
can point along a single direction. For instance, in the case of the
$B_{1g}^{-}$ AM phase shown in Fig. \ref{fig:spin_texture}, $\mathbf{d}\left(\mathbf{k}\right)=k_{x}k_{y}\hat{\mathbf{z}}$
along the entire $k_{z}=0$ plane, where $\langle S_z \rangle = \pm \hbar/2 $; interestingly, this would correspond to the $\alpha$-phase of the nematic-spin-nematic state introduced in Ref. \citep{Fradkin2007}. Conversely, in the $B_{1g}^{-}$ AM phase in a hexagonal system, $\mathbf{d}\left(\mathbf{k}\right)$ at $k_z = 0$ points in-plane and has the same form as in the $\beta$-phase of Ref. \citep{Fradkin2007}. We also note that, in a crystalline environment, magnetic
multipolar moments of different ranks $l$ are mixed. For example, in the case
of $B_{1g}^{-}$ or $B_{2g}^{-}$ AM order parameters in the hexagonal
group, while $\left(d_{x}\sigma^{x}+d_{y}\sigma^{y}\right)$ corresponds
to a magnetic octupolar moment and $\left(d_{x}\left(\mathbf{k}\right),\,d_{y}\left(\mathbf{k}\right)\right)$
is a $d$-wave form factor, $d_{z}\sigma^{z}$ corresponds to a magnetic
hexadecapolar toroidal moment and $d_{z}\left(\mathbf{k}\right)$
is a $g$-wave form factor (using the notation of Ref. \citep{Kusunose2018}). 

In real space,
the non-collinearity of the AM state is associated with the fact that different magnetic Wyckoff sites allow different magnetic moment orientations and that there will always be a magnetic Wyckoff site with non-collinear moments in altermagnets. We illustrate
it in Fig. \ref{fig:Materials}(b) for the proposed AM phase of MnF$_{2}$. Here, the magnetic symmetries of the Wyckoff sites occupied by the Mn and F atoms, labeled respectively 2a (purple) and 4f (grey), force the moments to point along the $c$ axis, as shown in Fig.~\ref{fig:Materials}(b). Meanwhile, for the Wyckoff sites labeled 8j (black), the moments point in the plane, as expected for a magnetic octupole moment (see Appendices \ref{sec_Abinitio_A}-\ref{sec_Abinitio_C} for more details).  Although
these Wyckoff sites are not occupied by atoms in MnF$_{2}$, the resulting non-collinearity of the spin-density will be manifested in the band structure in momentum space. Moreover, in other AM materials, the Wyckoff sites with non-collinear moments is not necessarily unoccupied. For instance, if the orthorhombic perovskite
CaMnO$_{3}$ were to realize the intra-unit-cell ``antiferromagnetic''
phase $G_aC_bA_c$ discussed above, which corresponds to an
$A_{1g}^{-}$ AM order parameter, Fig.~\ref{fig:Materials}(a), obtained from DFT calculations, shows that every
atom in the unit cell displays non-coplanar magnetic moments.

We end this section by noting an analogy between AM in the presence of SOC and unconventional superconductors. It has been pointed out the similarity between the d-wave Zeeman-splitting of an altermagnet with the gap function of a singlet d-wave superconductor \cite{Smejkal2022_1}. On the other hand, Eq. (\ref{eq:H_int}) resembles the Hamiltonian of a \emph{triplet} superconductor:

\begin{equation}
\mathcal{H}_{\mathrm{SC}}=-\sum_{\mathbf{k}}\Delta  \left[\mathbf{d}_\mathrm{SC}\left(\mathbf{k}\right)\cdot\boldsymbol{\sigma} \right]_{ss'} c_{\mathbf{k}s}^{\dagger} \left(i\sigma_y c_{\mathbf{-k}}^{\dagger}\right)_{s'} + \mathrm{h.c.}\,\label{eq:H_SC}
\end{equation}

The main difference is in the structures of the particle-hole and particle-particle condensates in spin space, which enforce the superconducting d-vector to be odd in momentum, $\mathbf{d}_\mathrm{SC}\left(-\mathbf{k}\right) = -\mathbf{d}_\mathrm{SC}\left(\mathbf{k}\right)$, whereas the altermagnetic d-vector must be even, $\mathbf{d}\left(-\mathbf{k}\right) = +\mathbf{d}\left(\mathbf{k}\right)$. Another consequence of this property is that the electronic spectrum of the superconductor consists of the sum in quadrature of the non-interacting and interacting coefficients:

\begin{equation}
E^{\mathrm{SC}}_{\pm}\left(\mathbf{k}\right)= \pm \sqrt{\varepsilon^2_{\mathbf{k}} + \left| \Delta \right|^2 \left| \mathbf{d}_\mathrm{SC}\left(\mathbf{k}\right)\right|^2}\, \ ,\label{eq:E_SC}
\end{equation}
whereas the electronic spectrum of the altermagnets consists of the sum of the two terms, as we will show in Eq. (\ref{eq:E}).

\section{Symmetry-protected nodal lines of the Zeeman-split bands}\label{sec:nodal_lines}

We are now in position to investigate the properties of the electronic
spectrum of an altermagnet in the presence of SOC. For simplicity, we first consider the case of
an Ising-like AM order. Using Eq. (\ref{eq:H_int}), diagonalization
of $\mathcal{H}=\mathcal{H}_{0}+\mathcal{H}_{\mathrm{int}}$ is straightforward
and yields two dispersions:
\begin{equation}
E_{\pm}\left(\mathbf{k}\right)=\varepsilon_{\mathbf{k}}\pm\lambda\,\Phi\left|\mathbf{d}\left(\mathbf{k}\right)\right|\ .\label{eq:E}
\end{equation}

\begin{figure}[t]
\centering
\centering \includegraphics[width=0.52\linewidth]{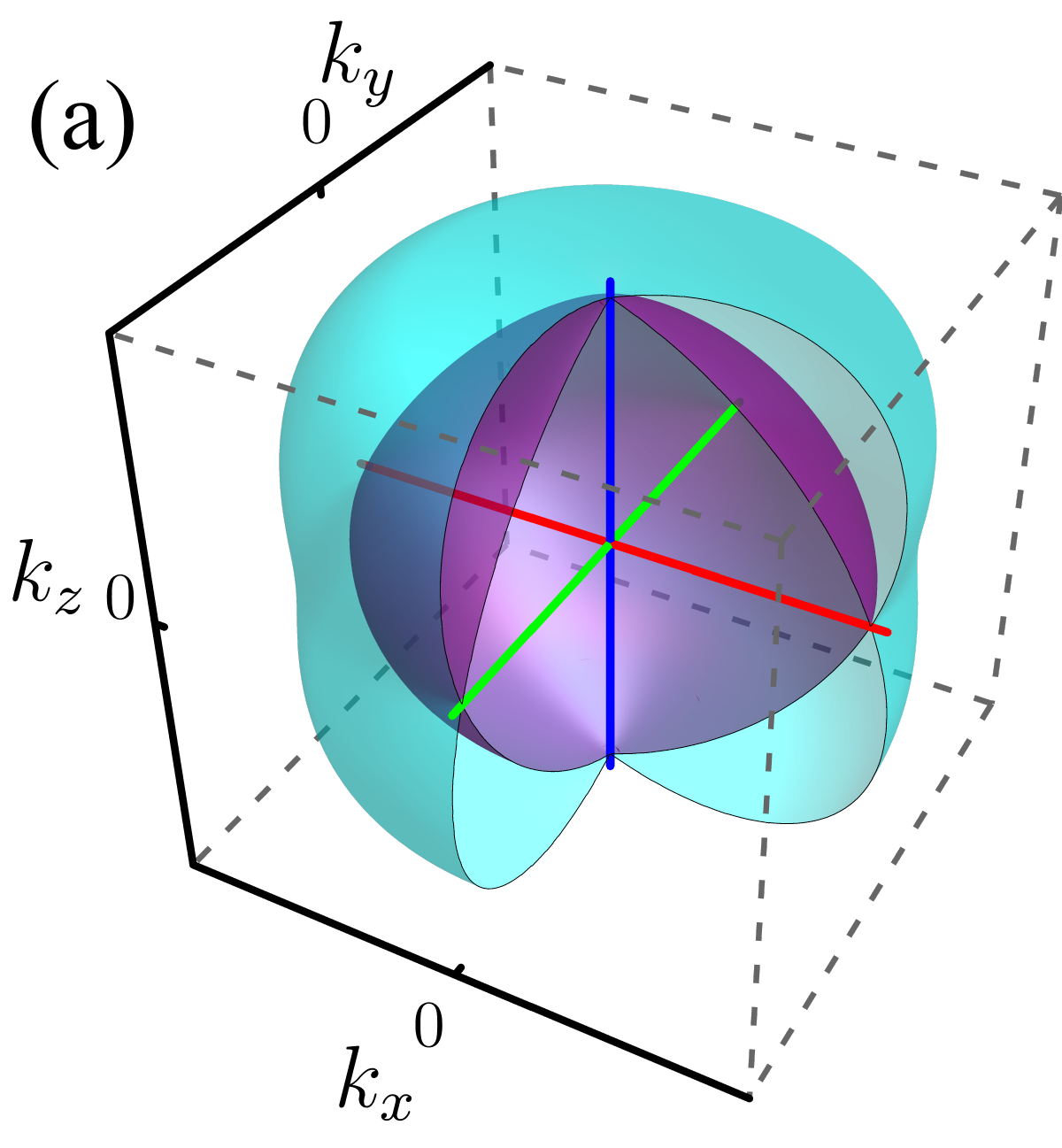} \hfil{}  \includegraphics[width=0.455\linewidth]{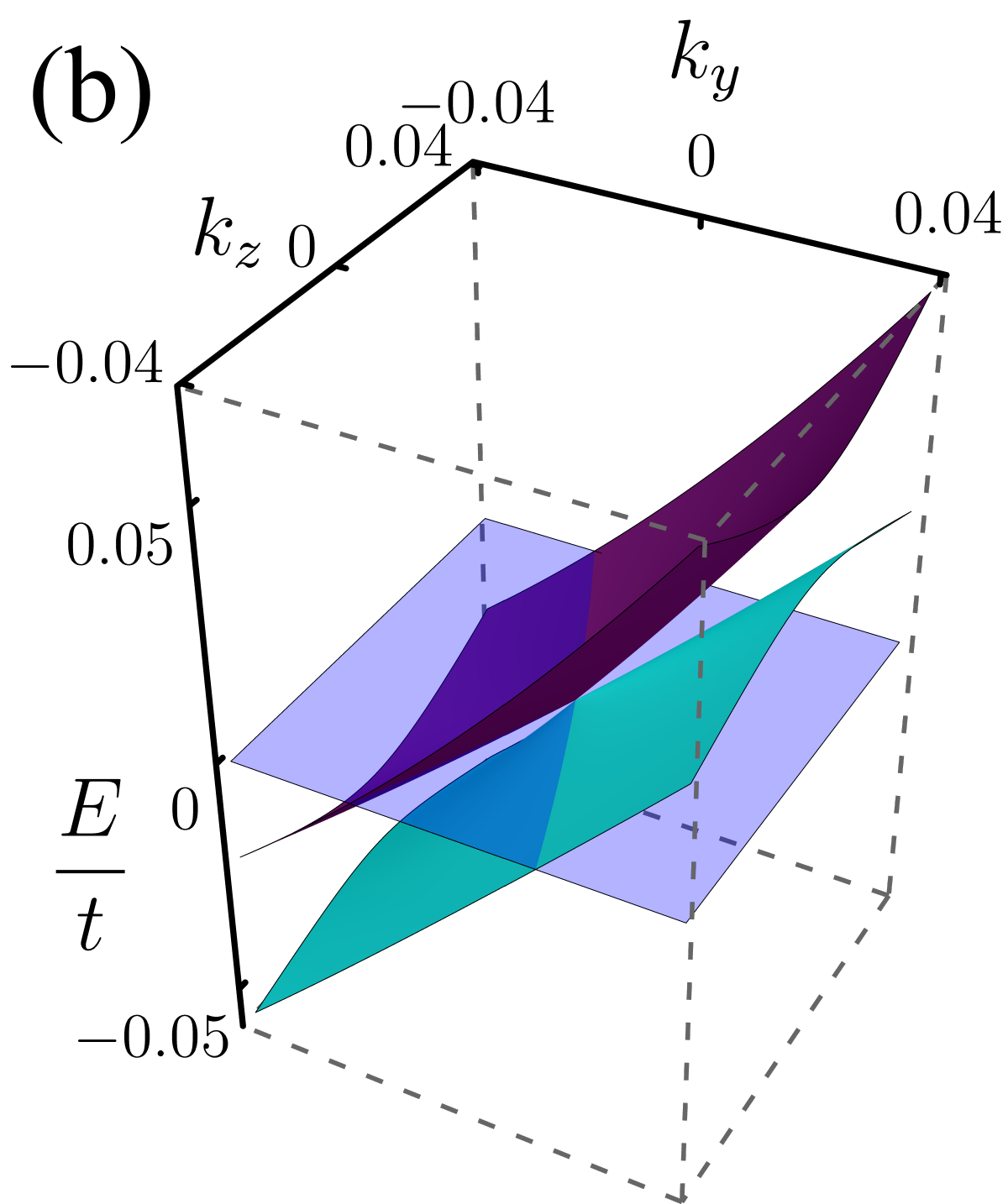} \vfil{}  \includegraphics[width=0.625\linewidth]{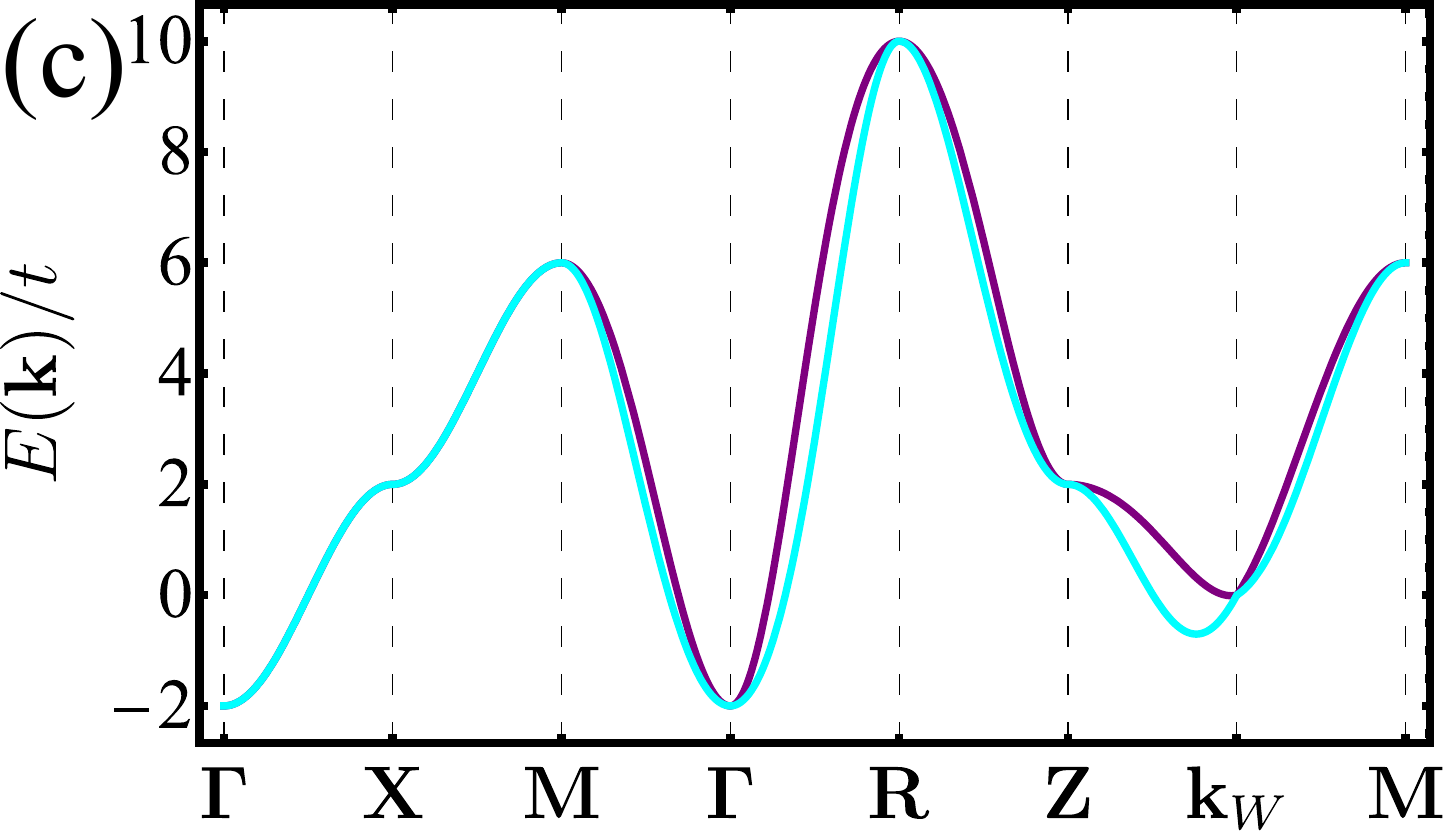}  \hfil{}   \includegraphics[width=0.35\linewidth]{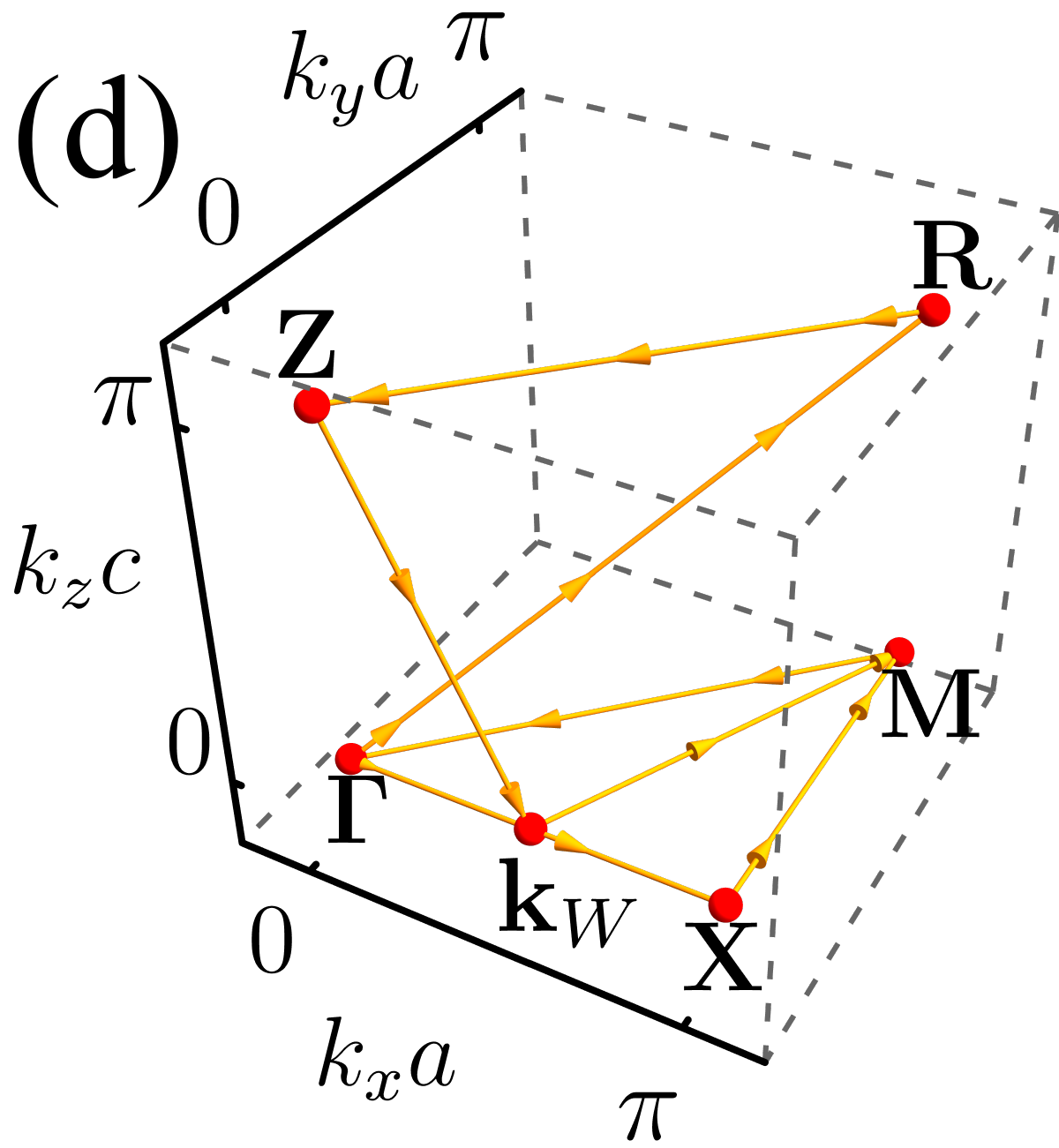} 
\caption{\textbf{Nodal lines and Weyl-like Fermi-surface pinch points.} (a) Spin-splitting of the Fermi surface of an altermagnetic metal whose AM order parameter $\Phi$ transforms as the tetragonal $B_{1g}^{-}$ irrep. Cyan and purple denote different Fermi surface sheets. Six pinch points where the Zeeman splitting vanishes emerge at the intersection with the nodal lines $\mathcal{L}_x$ (red), $\mathcal{L}_y$ (green), and $\mathcal{L}_z$ (blue). For better visualization, the Fermi-surface sheets for $k_x > 0$ and $k_y < 0$ are not shown. (b) The electronic dispersion in the vicinity of the pinch point has the shape of a tilted cone, characteristic of a type-II Weyl node. (c) Zeeman splitting of a simple nearest-neighbor tight-binding dispersion $E(\mathbf{k})$ (in units of the electronic hopping $t$) in the AM state along the high-symmetry directions of the Brillouin zone shown in panel (d). The Weyl-like pinch point is located in this case at $\mathbf{k}_W$.}\label{Fig_02}
\end{figure}

In agreement with previous works \citep{Kusunose2019,Smejkal2022_2,Spaldin2022,Zunger2020,Winkler2023},
we find a momentum-dependent Zeeman splitting of the bands, $\Delta E\left(\mathbf{k}\right)\equiv E_{+}\left(\mathbf{k}\right)-E_{-}\left(\mathbf{k}\right)$,
which in our case is given by $\Delta E\left(\mathbf{k}\right)\propto\left|\mathbf{d}\left(\mathbf{k}\right)\right|$.
Therefore, the Zeeman splitting only vanishes when the three components
of $\mathbf{d}\left(\mathbf{k}\right)$ are simultaneously zero. Because
of the symmetry properties of $\mathbf{d}\left(\mathbf{k}\right)$,
this condition is not as restrictive as it may seem. Indeed, for all
cases shown in Table \ref{tab:classification}, the vector $\mathbf{d}\left(\mathbf{k}\right)$
vanishes along either lines (for 1D irreps) or planes (for some multi-dimensional
irreps) in momentum space, giving rise to nodal lines/planes in the
band structure.  In contrast, in the case where rotations in spin-space are decoupled from real space operations (i.e. in the absence of SOC), $\mathbf{d}$ has effectively only one component. As a result, it vanishes along planes and the Zeeman splitting is characterized by nodal planes, as discussed in Ref. \cite{Mazin2021}.

To shed further light on the nature of the Zeeman splitting nodes, we consider
the specific case of an AM order parameter $\Phi$ that transforms
as $B_{1g}^{-}$ in a tetragonal system. It follows from Table \ref{tab:classification}
that $\mathbf{d}\left(\mathbf{k}\right)=0$ defines three nodal lines
($\mathcal{L}_{x}$, $\mathcal{L}_{y}$, $\mathcal{L}_{z}$) determined
by the intersection between the three high-symmetry planes $k_{x}=0$,
$k_{y}=0$, and $k_{z}=0$, such that $\mathcal{L}_{x}$ corresponds
to $k_{y}=k_{z}=0$, $\mathcal{L}_{y}$ to $k_{x}=k_{z}=0$ and $\mathcal{L}_{z}$
to $k_{x}=k_{y}=0$ (see also Ref. \cite{Mazin2021}). Along these lines, displayed respectively as red, green, and blue lines in Fig. \ref{Fig_02}(a), the Zeeman splitting of the band structure vanishes, as shown in Fig. \ref{Fig_02}(c) for the case of a simple nearest-neighbors tight-binding model (in this and the remainder figures, we set $\eta = 1/2$ for concreteness).

The topological properties of the nodal lines $\mathcal{L}_{\alpha}$ can
be obtained from the symmetry properties of $\mathcal{H}_{\mathrm{int}}$
under time-reversal, charge-conjugation, and chiral operations.
Following the tenfold classification of Ref. \citep{Schnyder2014},
we conclude that the nodal lines belong to class C of gapless topological
phases, as they correspond to ``Fermi surfaces\textquotedblright{}
of codimension $p=2$ that lie along high-symmetry directions. Thus,
the nodal lines are topologically trivial with respect to non-crystalline symmetries.
Nevertheless, as shown in Ref. \citep{Schnyder2014} (see also Ref.
\citep{Timm2022}), topologically trivial lines can still be protected
by crystalline symmetries that leave $\mathcal{H}_{\mathrm{int}}$
invariant. This is precisely the case for the nodal lines $\mathcal{L}_{\alpha}$:
the two planes along which a given nodal line lies are actually vertical
or horizontal mirror planes of the point group $D_{4h}$. Therefore,
the AM nodal lines are symmetry-protected. Importantly, this conclusion
holds for all AM orders in Table \ref{tab:classification} that transform
as a 1D irrep: as shown in Table \ref{Tab_Nodal} in Appendix \ref{sec_NodalLines}, at least one
of the planes along which a given nodal line lies is a crystallographic
mirror plane. The case of multi-dimensional irreps is more subtle,
as nodal planes also emerge for certain order-parameter configurations
(see Appendix \ref{sec_NodalLines}). 

The presence of symmetry-protected nodal lines in the AM state has
important consequences for its low-energy electronic properties, particularly
in the case of a metallic system. To show this, we once again focus
on the case of a $B_{1g}^{-}$ AM order parameter $\Phi$ in a tetragonal
system with a generic parabolic dispersion $\varepsilon_{\mathbf{k}}=\frac{k^{2}}{2m}-\mu$.
As shown in Fig. \ref{Fig_02}(a), the crossings between the three nodal lines
$\left\{ \mathcal{L}_{x},\mathcal{L}_{y},\mathcal{L}_{z}\right\} $
and the Fermi surface defines six pinch points where the Fermi surface
is not Zeeman-split, $\mathbf{k}_{1\pm}=\left(\pm k_{F},0,0\right)$,
$\mathbf{k}_{2\pm}=\left(0,\pm k_{F},0\right)$ and $\mathbf{k}_{3\pm}=\left(0,0,\pm k_{F}\right)$,
with $k_{F}=\sqrt{2m\mu}$. Writing the Hamiltonian as $\mathcal{H}=\sum_{k}c_{\mathbf{k}s}^{\dagger}H_{\mathbf{k}}^{ss'}c_{\mathbf{k}s'}^{\phantom{\dagger}}$
, expanding around $k_{1+}$, and defining $\mathbf{q}\equiv\mathbf{k}-\mathbf{k}_{1+}$,
we obtain:
\begin{equation}
H_{\mathbf{q}}=v_{F}q_{x}\sigma^{0}+\lambda\Phi k_{F}\left(q_{z}\sigma^{y}+\eta q_{y}\sigma^{z}\right)\ ,\label{eq:Weyl}
\end{equation}
with $v_{F}=k_{F}/m$. Using the results of Ref. \citep{Bernevig2015},
we identify $H_{\mathbf{q}}$ as the effective Hamiltonian of a type-II Weyl
node. The dispersion around the pinch point, shown explicitly in Fig.
\ref{Fig_02}(b), displays the characteristic tilted-cone structure of type-II Weyl points.
The topological nature of the pinch points is further confirmed by
a straightforward calculation of the Berry phase, which yields $\gamma=\pm\pi$
(see Appendix \ref{sec_NodalLines}). Beyond this specific case,
we find that behaviors analogous to type-II Weyl nodes emerge at the intersections between
the nodal lines and the Fermi surface for all 1D-irrep AM $\Phi$
order parameters of Table \ref{tab:classification}. In the particular case of the
$B_{1g}^{-}$ and $B_{2g}^{-}$ irreps of $D_{6h}$, these Weyl-like points have a Berry phase of $\gamma=\pm2\pi$, and thus behave as double Weyl points \cite{Bernevig2012} (see Appendix \ref{sec_NodalLines}). Note that these Weyl-like nodes are qualitatively different from actual Weyl points that can emerge in certain magnetically ordered states \cite{Zhou2023}. We emphasize that our main
conclusions for the band structure of AM systems -- i.e. the existence
of symmetry-protected nodal lines -- are enabled
by the multi-component nature of the vector $\mathbf{d}\left(\mathbf{k}\right)$, regardless of the magnitude of the anisotropy parameter
$\eta$ in Table \ref{tab:classification}.

\begin{figure*}[t]
\centering
\centering \includegraphics[width=0.32\linewidth]{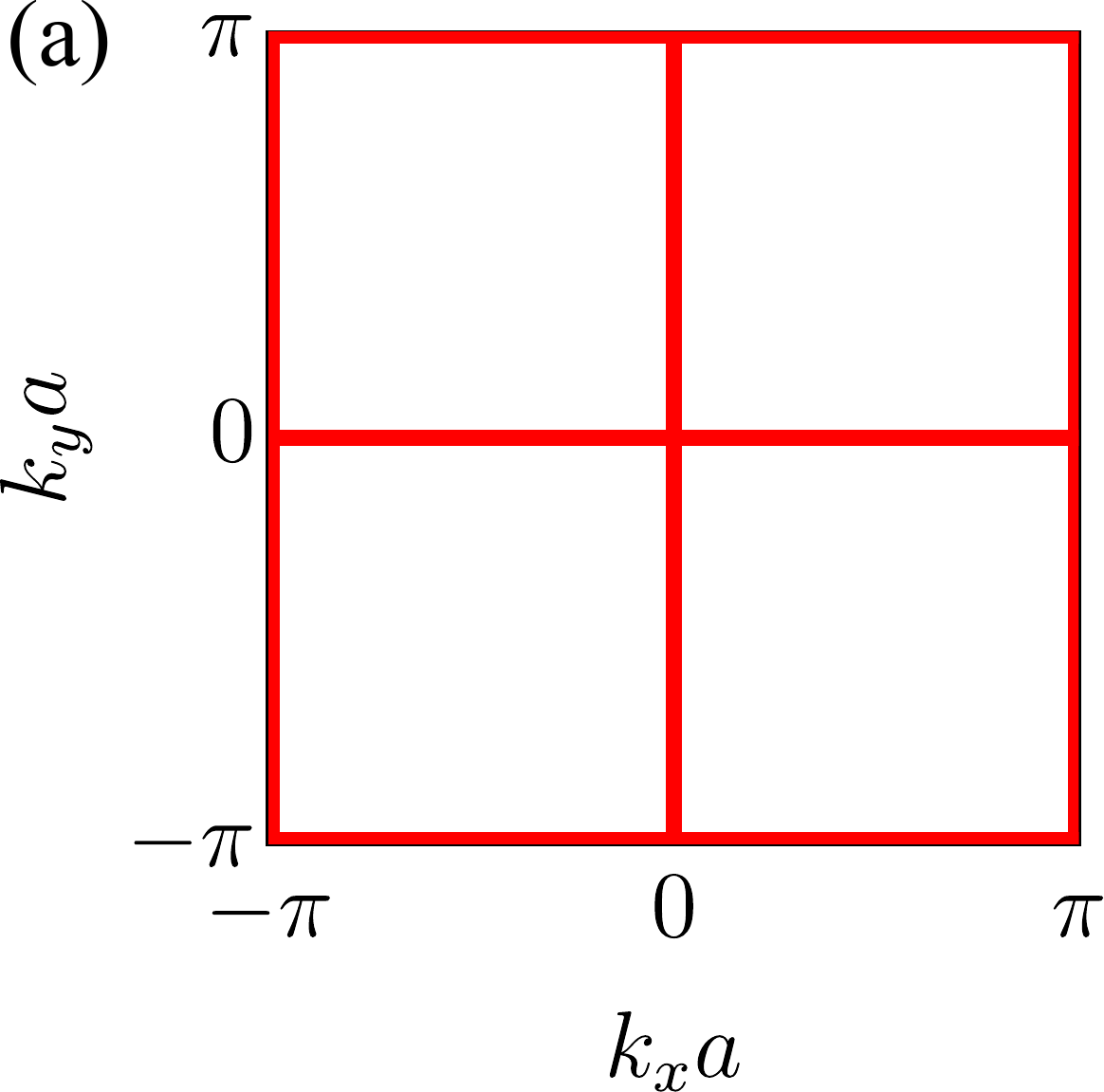} \hfil{}  \includegraphics[width=0.32\linewidth]{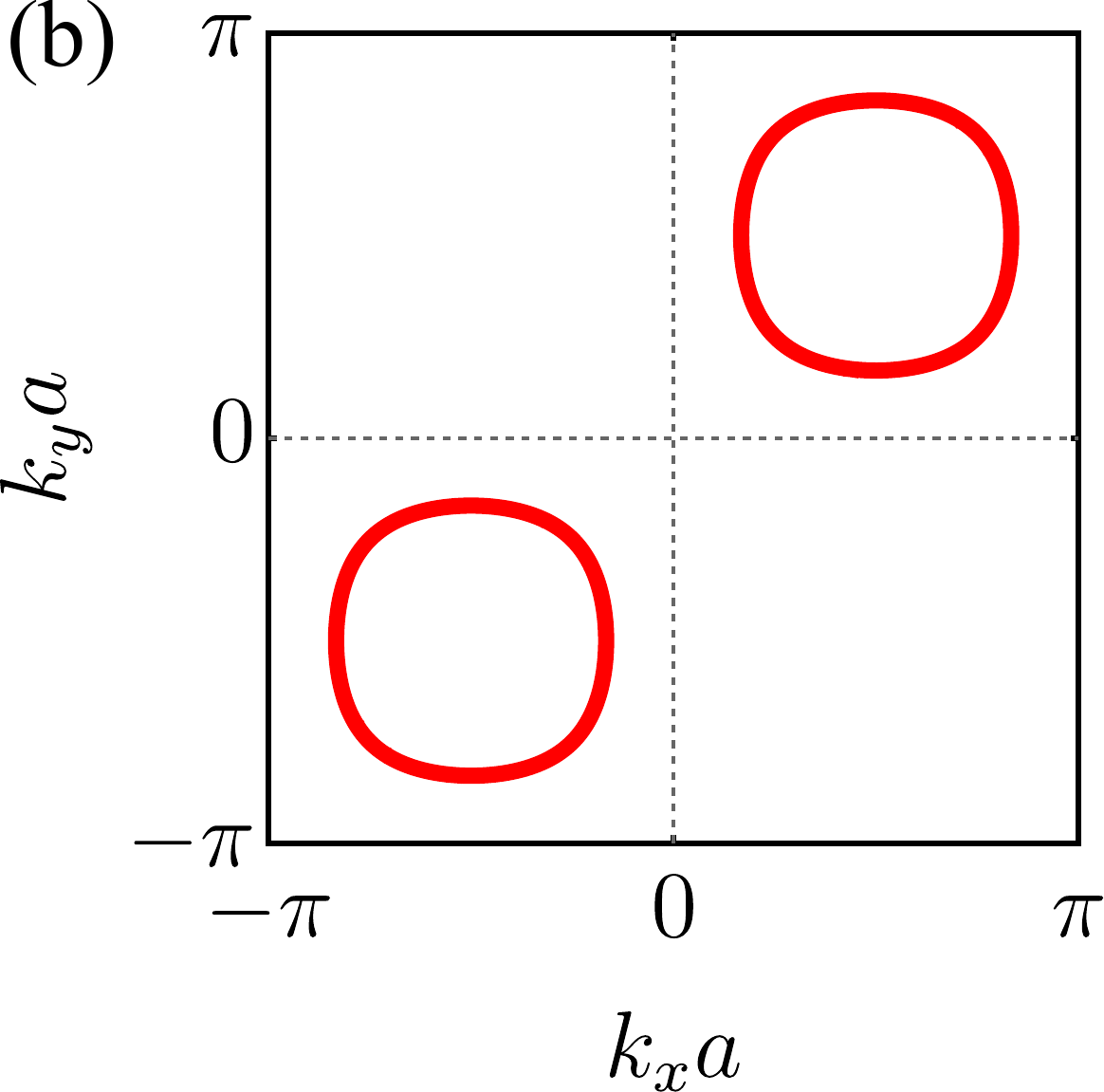} \hfil{}  \includegraphics[width=0.32\linewidth]{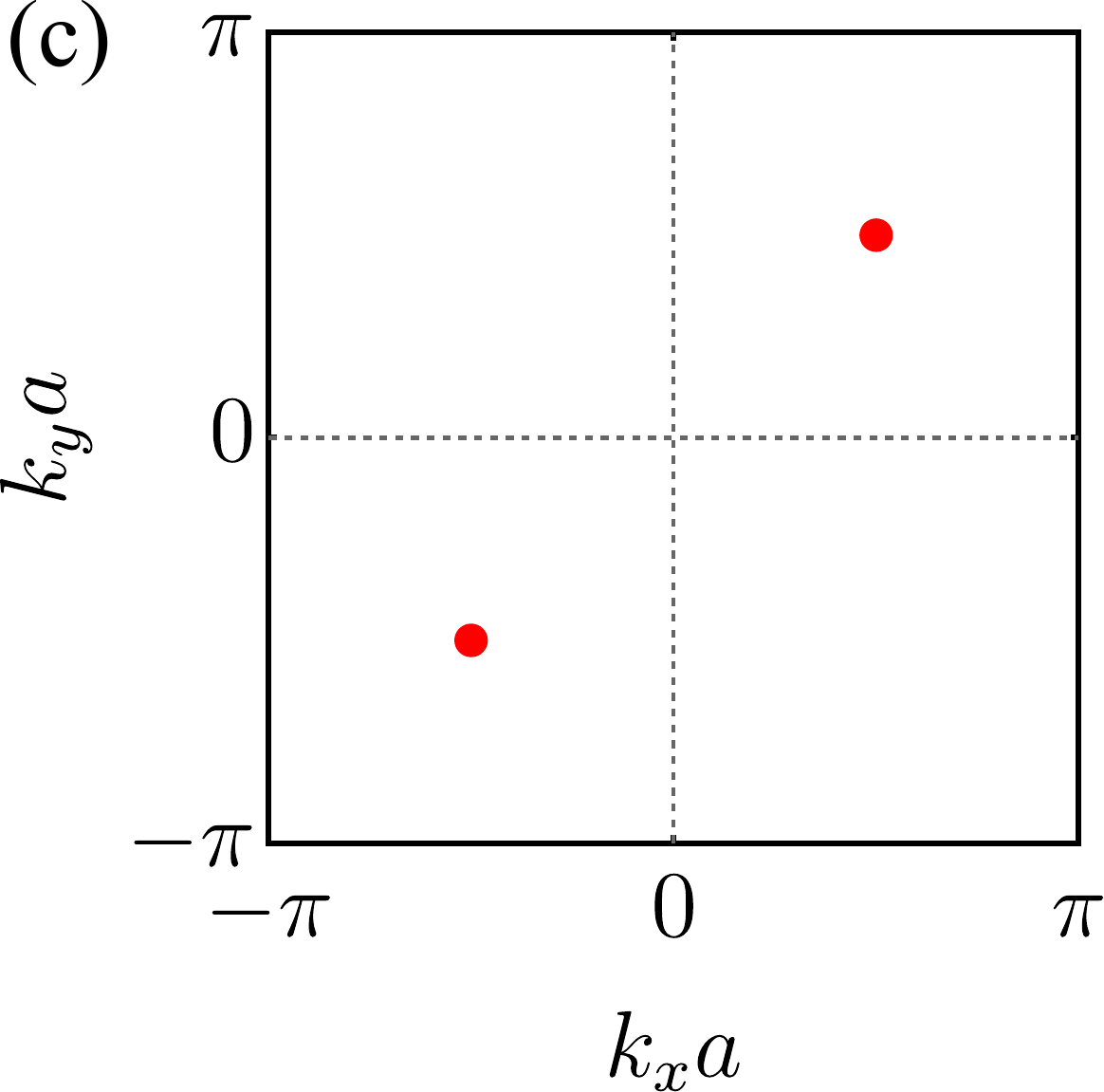} \vfil{} \includegraphics[width=0.32\linewidth]{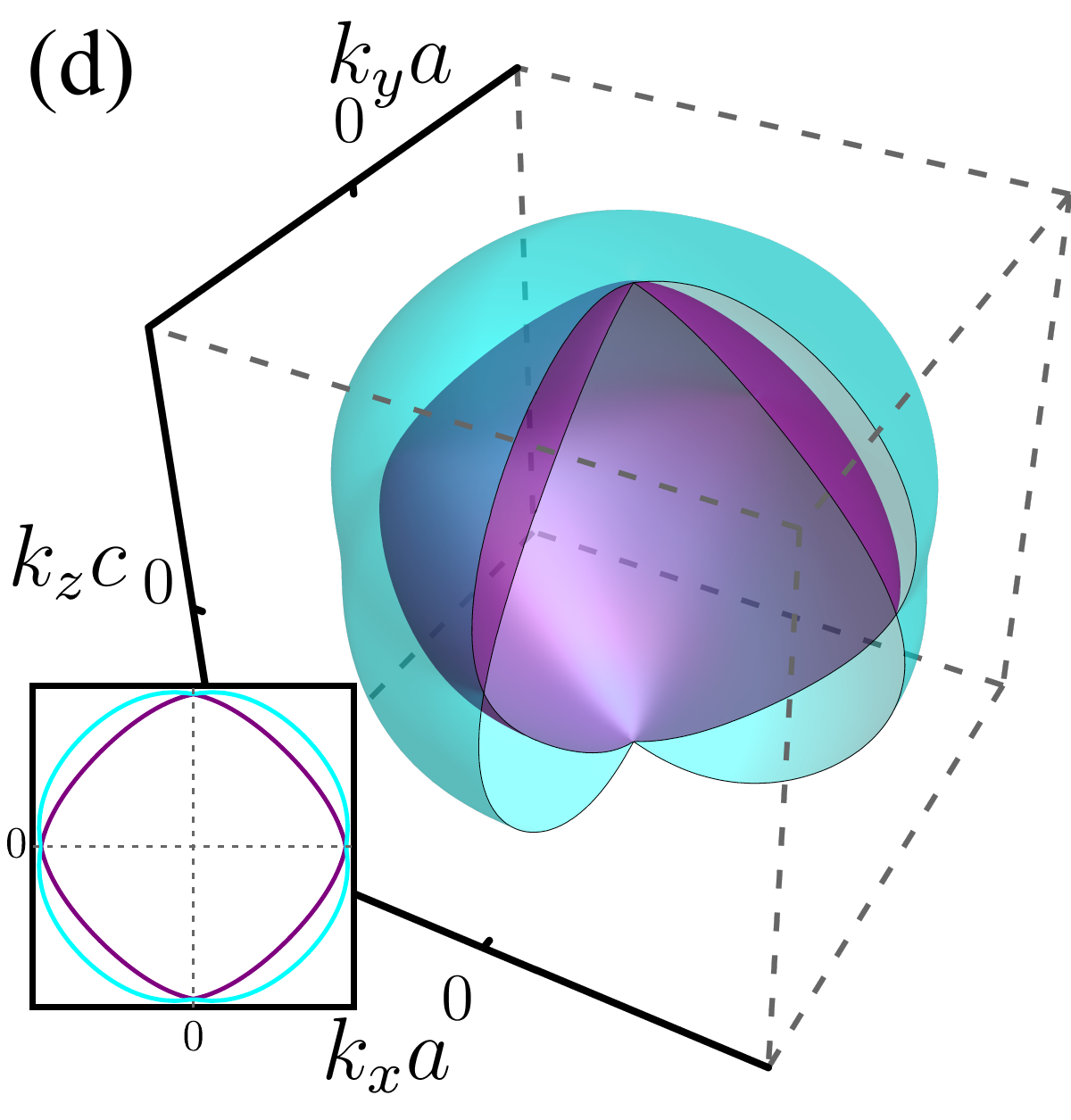} \hfil{}  \includegraphics[width=0.32\linewidth]{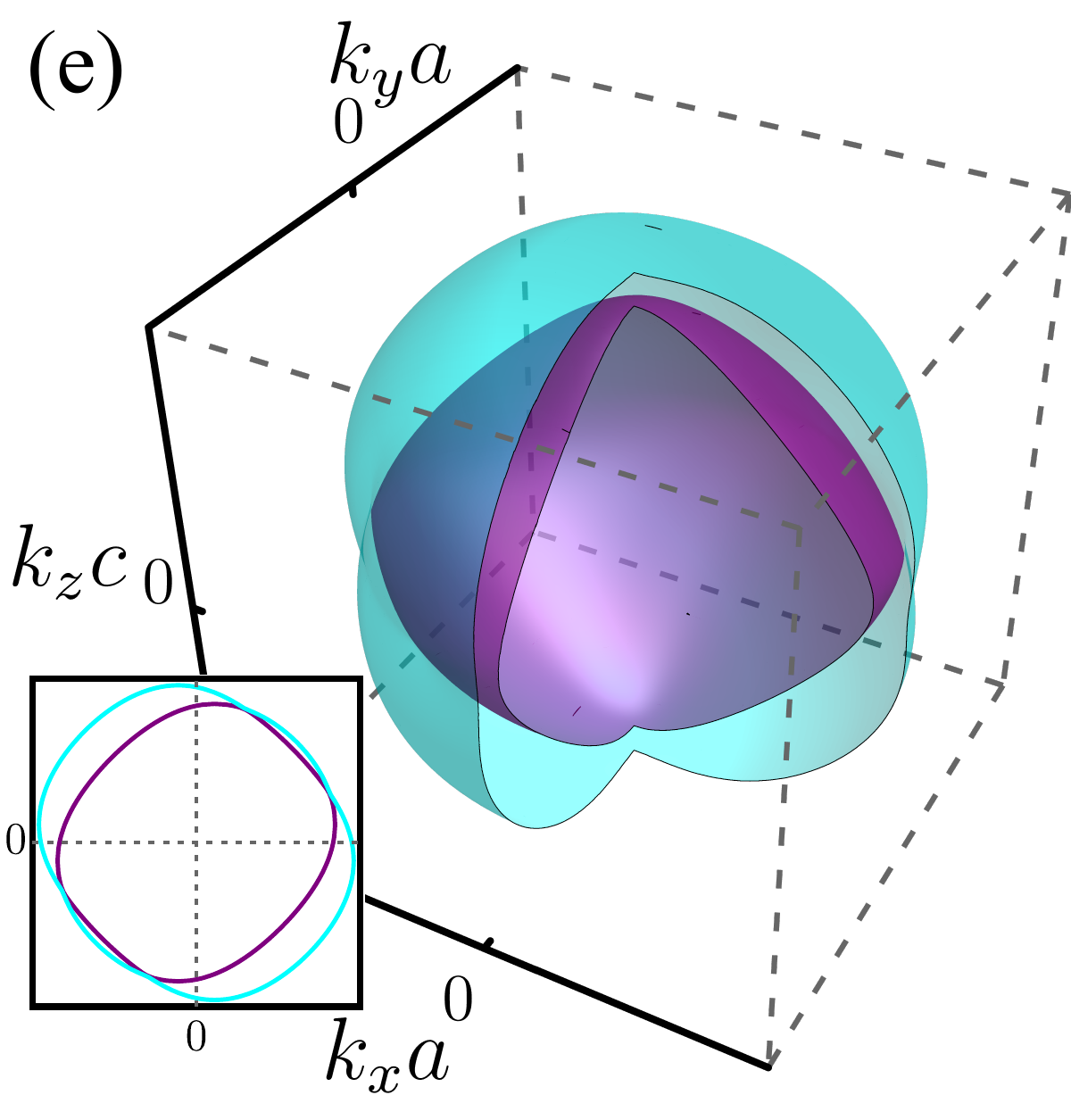} \hfil{}  \includegraphics[width=0.32\linewidth]{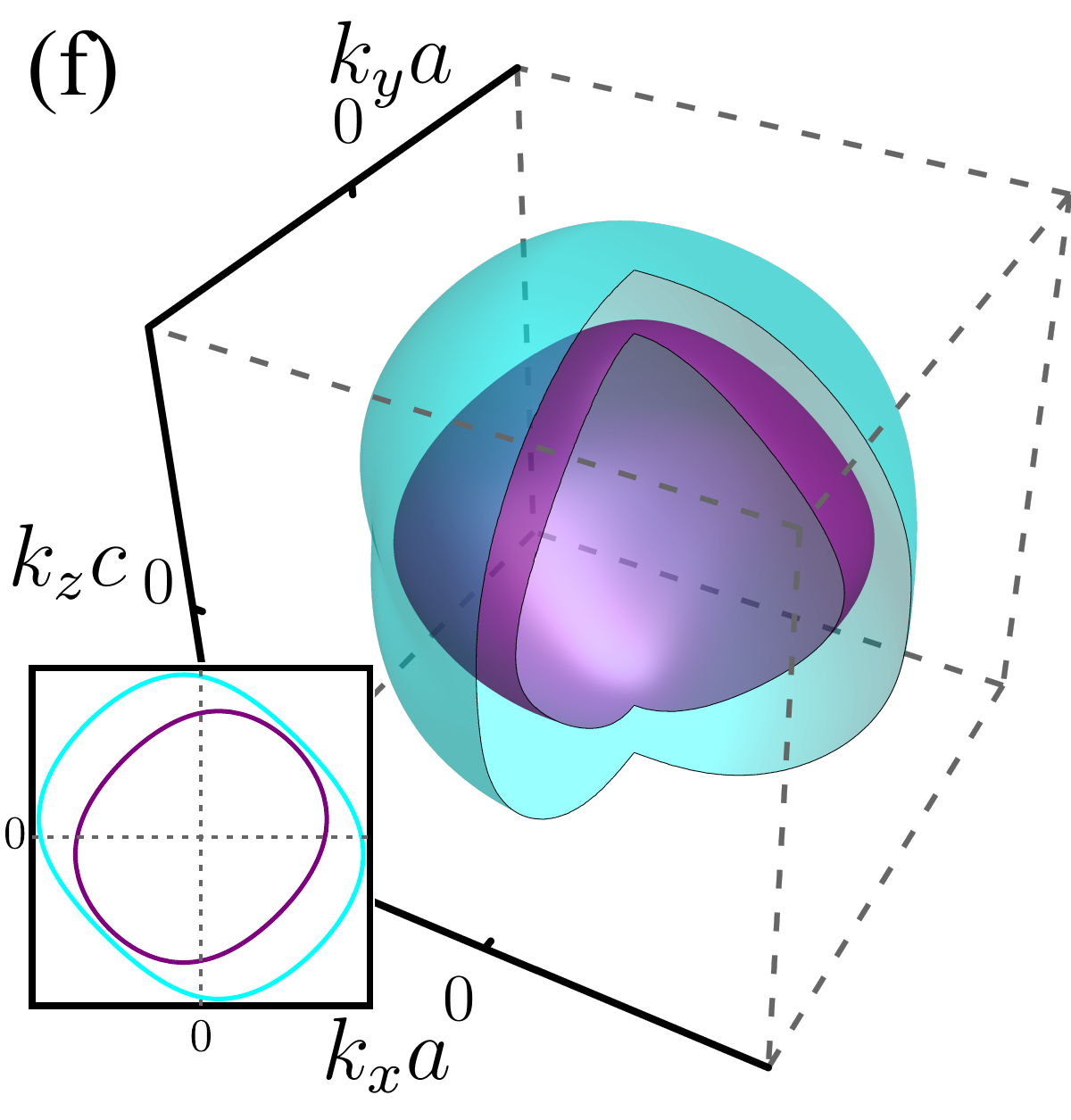}
\caption{\textbf{Impact of a magnetic field on the nodal lines and Weyl-like Fermi-surface pinch points.} In this figure, a magnetic field $\mathbf{h}$ is applied along the $z$-axis of a tetragonal system with $B^{-}_{1g}$ AM order. Panels (a)-(c) show the evolution of the nodal lines (red) on the $k_z = 0$ plane of the first Brillouin zone for $h=0$, $h=h_c/2$, and $h= h_c \equiv \frac{\lambda \Phi \eta}{g_s \mu_B}$, respectively. As the field increases, the nodal lines form loops that collapse onto points at a critical value of the field. Panels (d)-(f) show the evolution of the split sheets of the Fermi surface for the same values of $h$; the insets show the Fermi surfaces projected onto the $k_z = 0$ plane. Note that the pinch points, which behave as Weyl nodes, annihilate for a critical field $h_c^* < h_c$, such that for $h=h_c$ [panel (f)], the Fermi surfaces are disconnected. A simple nearest-neighbors tight-binding model was used to parametrize the band dispersion.
}\label{Fig_LineFermi}
\end{figure*}

\begin{figure}[t]
\centering
\centering \includegraphics[width=0.9\linewidth]{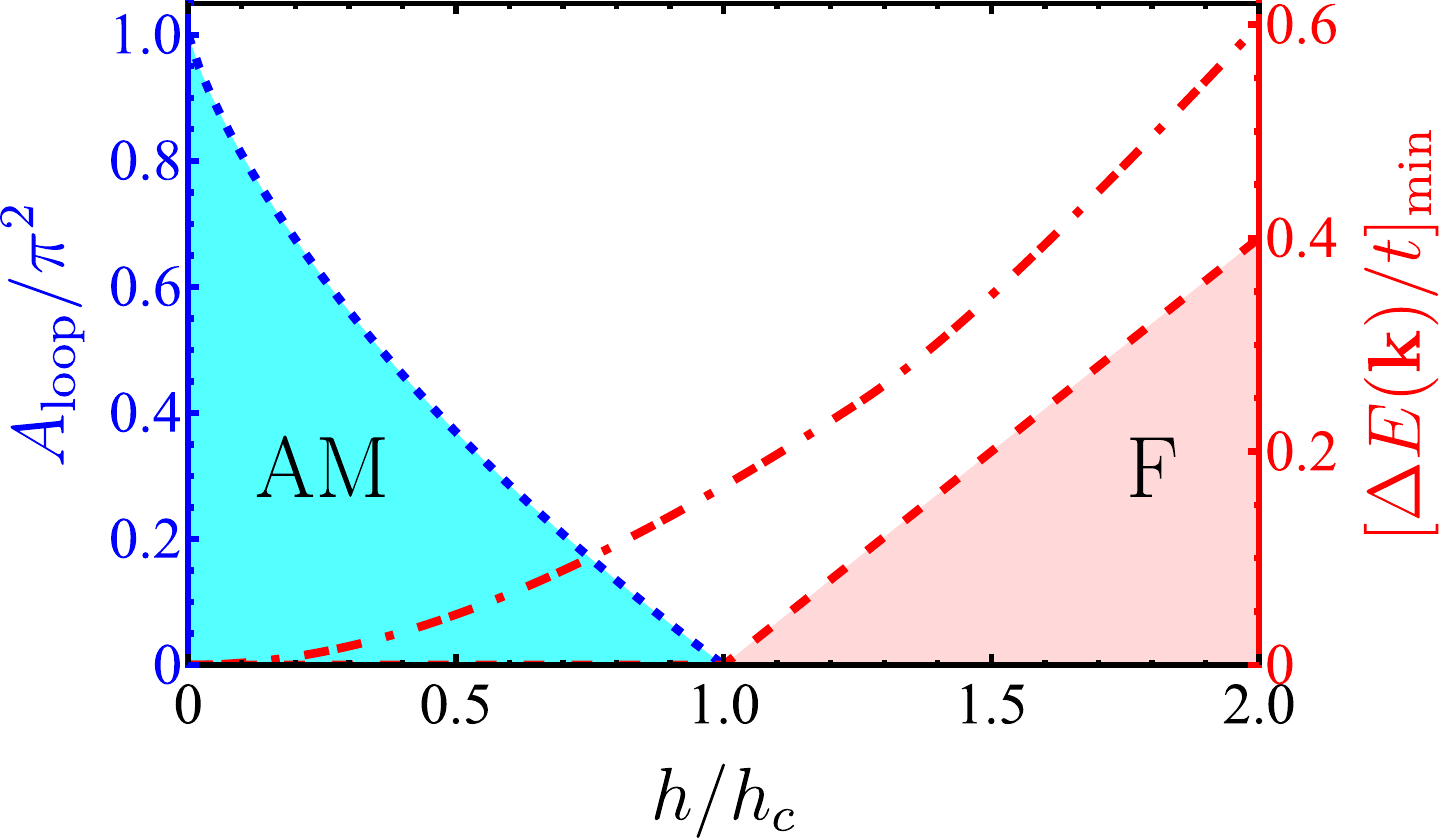}
\caption{\textbf{Field-driven altermagnetic (AM) to ferromagnetic (F) transition.} The ``phase diagram" describes the transition between the tetragonal $B^{-}_{1g}$ AM state, characterized by a nodal Zeeman splitting, and the F state, characterized by a nodeless Zeeman splitting. Here, $A_{\mathrm{loop}}$ (dashed blue line) refers to the area of the closed nodal lines on the $k_z = 0$ plane (see Fig. \ref{Fig_LineFermi}(b)), whereas $[\Delta E(\mathbf{k})]_{\mathrm{min}}$ denotes the minimum value of the Zeeman splitting between the bands. When $\mathbf{h}$ is along the $z$-axis, the AM-F transition is topological and occurs at the critical field $h_c = \frac{\lambda \Phi \eta}{g_s \mu_B}$ (dashed red line). For $\mathbf{h}$ along the main diagonal of the $k_z = 0$ plane, the AM-F transition is trivial and takes place for an infinitesimal field (dash-dotted red line).}\label{Fig_PhasDiag}
\end{figure}

\section{Topological AM-F transition induced by an external magnetic field}\label{sec:topo_transition}

The existence of symmetry-protected nodal lines enabled by the SOC fundamentally impacts
the response of an AM phase to a magnetic field. In the presence of
a field $\mathbf{h}$, the Hamiltonian acquires a Zeeman term of the
form $\mathcal{H}_{\mathrm{Z}}=g_{s}\mu_{B}\sum_{\mathbf{k}}\mathbf{h}\cdot\boldsymbol{\sigma}_{ss'}c_{\mathbf{k}s}^{\dagger}c_{\mathbf{k}s'}^{\phantom{\dagger}}$,
where $g_{s}$ denotes the effective g-factor and $\mu_{B}$, the
Bohr magneton. While in a paramagnet this term leads to a uniform Zeeman splitting
of the electronic bands, $\Delta E_\mathrm{Z} = g_{s}\mu_{B}h$, the situation
in an altermagnet is qualitatively different. To see this, we note
that $\mathcal{H}_{\mathrm{Z}}$ can be incorporated into $\mathcal{H}_{\mathrm{int}}$
in Eq. (\ref{eq:H_int}) provided that the vector $\mathbf{d}\left(\mathbf{k}\right)$
is replaced by:
\begin{equation}
\tilde{\mathbf{d}}\mathbf{\left(\mathbf{k}\right)}=\mathbf{\mathbf{d}\left(\mathbf{k}\right)}-\frac{g_{s}\mu_{B}}{\lambda\Phi}\,\mathbf{h}\ .\label{eq:field}
\end{equation}

The key point is that if the direction of the external field $\mathbf{h}$
is perpendicular to one of the mirror planes protecting the AM nodal lines,
the new $\mathcal{H}_{\mathrm{int}}$ will remain invariant under
reflection with respect to the corresponding mirror. Consequently,
the nodal lines that lie on that mirror plane will remain symmetry-protected
even in the presence of $\mathbf{h}$. While this symmetry alone is
not enough to ensure that a nodal line must exist, it does imply that
if the nodal line exists before the application of the field, an infinitesimal
field will not be able to gap it out, but only to move it along the
mirror plane. Upon increasing the magnetic field, the nodal lines
form closed loops that eventually collapse for a critical field value
$h_{c}$, whose expression we give below. Thus, the momentum-dependent Zeeman splitting of the band
structure, $\Delta E\left(\mathbf{k}\right)$, is nodal for $h<h_{c}$
but nodeless for $h>h_{c}$. Since a nodal Zeeman splitting is a typical
feature of an altermagnet, whereas a nodeless Zeeman splitting is characteristic
of a ferromagnet, we denote the topological nodal-to-nodeless transition at $h_{c}$
an AM-F transition.

It is instructive to illustrate these results in the specific case
of a tetragonal $B_{1g}^{-}$ AM order parameter. Consider first a
field $\mathbf{h}$ applied along $\hat{\mathbf{z}}$, which is perpendicular
to the mirror plane where the zero-field nodal lines $\mathcal{L}_{x}$
($k_{y}=k_{z}=0$) and $\mathcal{L}_{y}$ ($k_{x}=k_{z}=0$) are located.
Solving $\left|\tilde{\mathbf{d}}\mathbf{\left(\mathbf{k}\right)}\right|=0$,
we find the condition $k_{z}=0$, $k_{x}k_{y}=\frac{g_{s}\mu_{B}}{\lambda\Phi\eta}\,h$,
which corresponds to two new nodal lines $\mathcal{L}'_{1}$ and $\mathcal{L}'_{2}$.
Note that the number of nodal lines decreases from $3$ to $2$ upon
application of an infinitesimal magnetic field, because line $\mathcal{L}_{z}$
is immediately gapped due to the fact that it does not lie on the
relevant mirror plane. Like $\mathcal{L}_{x}$ and $\mathcal{L}_{y}$,
$\mathcal{L}'_{1}$ and $\mathcal{L}'_{2}$ are also located on the
$k_{z}=0$ plane; however, in contrast to the former, which are straight
nodal lines, the latter are hyperbolic nodal lines. In the continuum,
increasing $h$ just moves the foci of the two hyperbolas to larger
values of momentum. However, on the lattice, the new nodal lines form
closed loops that collapse at a critical magnetic field value $h_{c}$.
To determine $h_{c}$, we rewrite $\tilde{\mathbf{d}}\mathbf{\left(\mathbf{k}\right)}$
in terms of lattice harmonics and solve once again $\left|\tilde{\mathbf{d}}\mathbf{\left(\mathbf{k}\right)}\right|=0$.
The new nodal lines $\mathcal{L}'_{1}$ and $\mathcal{L}'_{2}$ are
now described by the equations $\sin\left(k_{z}c\right)=0$ and $\sin\left(k_{x}a\right)\sin\left(k_{y}a\right)=\frac{g_{s}\mu_{B}}{\lambda\Phi\eta}\,h$.
As shown in Figs. \ref{Fig_LineFermi}(a)-(c), they describe nodal loops that collapse
onto the points $\mathbf{k}_{c}=\pm\left(\frac{\mathrm{sgn}\left(h\right)\pi}{2a},\,\frac{\pi}{2a},\,0\right)$ at a topological transition
taking place at the critical value of the field
\be
h_{c}\equiv\frac{\lambda\Phi\eta}{g_{s}\mu_{B}}\, .
\ee
Note that
the location of the nodal loops on the $k_{z}=0$ plane -- first
and third quadrants for $h>0$ or second and fourth quadrants for
$h<0$ -- breaks the tetragonal symmetry of the lattice down to orthorhombic
by making the two in-plane diagonals inequivalent. This is a direct
consequence of the fact that there is a trilinear term in the free
energy coupling the $B_{1g}^{-}$ AM order parameter $\Phi$, the
out-of-plane magnetic field $h$ (which transforms as $A_{2g}^{-}$),
and the shear strain $\varepsilon_{xy}$ (which transforms as $B_{2g}^{+}$;
see also Refs. \citep{Fradkin2007,Patri2019,Steward2023}).

Not surprisingly, the evolution of the Weyl-like pinch points as
a function of an applied magnetic field mirrors that of the nodal
lines. Figs. \ref{Fig_LineFermi}(d)-(f) show the Zeeman-split Fermi surface in the
tetragonal $B_{1g}^{-}$ AM ordered phase for different values of
a field parallel to the $\hat{\mathbf{z}}$ axis; the insets show the Fermi surfaces projected onto $k_z  = 0$. For $h<h_{c}^{*}$,
the Fermi surface displays two pairs of Weyl nodes with opposite
Berry phases located at $\mathbf{k}'_{1\pm}=\pm k_{F}\left(\kappa_+,\kappa_-, 0\right)$
and $\mathbf{k}'_{2\pm}=\pm k_{F}\left(\kappa_-,\kappa_+,0\right)$, with $\kappa_\pm \equiv \mathrm{sgn}(h)^{\frac{1\pm1}{2}} \sqrt{1 \pm \sqrt{1-\alpha^2} }/\sqrt{2}$ and $\alpha \equiv \frac{2 g_s \mu_B h}{\lambda \Phi \eta k_F^2}$. At $h=h_{c}^{*}$, set by the condition $\alpha = 1$,
the pairs of Weyl points $\mathbf{k}'_{1+}$ and $\mathbf{k}'_{2+}$, as well as $\mathbf{k}'_{1-}$ and $\mathbf{k}'_{2-}$, annihilate, resulting in fully split
Fermi surfaces. Note that $h_{c}^{*}\leq h_{c}$, since the nodal
lines will generally cease to cross the Fermi surface before they
collapse. This behavior is analogous to the annihilation of nodes
in superconductors undergoing a nodal-to-nodeless transition \citep{Stanev2011,Fernandes2011,Khodas2012,Mazidian2013,Chubukov2016}. The important difference is that, while in the superconducting case the gap function has a single angular dependence, in Eq. (\ref{eq:E}) the effective Zeeman-gap $\left|\mathbf{d}\left(\mathbf{k}\right)\right|$ has three independent components that must simultaneously vanish to produce nodes. 

In the case of a field applied along either the $\hat{\mathbf{x}}$
or $\hat{\mathbf{y}}$ axes, the analysis above remains almost unchanged,
since these directions are also perpendicular to mirror planes on
which two nodal lines reside, resulting in a topological nodal-to-nodeless transition. The main differences are: the value of
the critical field, $h_{c}^{(x)}=h_{c}^{(y)}=h_{c}^{(z)}/\eta$, which
reflects the magnetic anisotropy of the tetragonal lattice encoded
by the parameter $\eta$; and the lower monoclinic symmetry of the
lattice, which reflects the triggering of an out-of-plane shear distortion
$\varepsilon_{xz}$/$\varepsilon_{yz}$ by the magnetic field. This analysis illustrates that the critical field $\mathbf{h}_c$ can be anisotropic and depends on parameters that vary among different materials.

The
situation changes substantially, however, when the field is not applied
along one of the three Cartesian axes. Consider, for example, a field
applied along the in-plane diagonal $\mathbf{h}=\frac{1}{\sqrt{2}}\left(h,h,0\right)$.
The condition $\left|\tilde{\mathbf{d}}\mathbf{\left(\mathbf{k}\right)}\right|=0$
now yields the system of equations $k_{y}k_{z}=\frac{g_{s}\mu_{B}}{\lambda\Phi\eta\sqrt{2}}\,h$,
$k_{x}k_{z}=\frac{g_{s}\mu_{B}}{\lambda\Phi\eta\sqrt{2}}\,h$, and
$k_{x}k_{y}=0$, which does not admit a solution. Consequently, an
infinitesimal field along this direction immediately gaps out all
nodal lines, rendering the AM-F transition trivial. Fig. \ref{Fig_PhasDiag} illustrates the field-driven AM-F ``phase diagram'', where the AM (F) phase is identified as that with nodal (nodeless) Zeeman splitting. The minimum Zeeman splitting, $[\Delta E(\mathbf{k})]_{\mathrm{min}}$, is only non-zero above the threshold $z$-axis field $h_c$, whereas the area of the nodal loops in Fig. \ref{Fig_LineFermi}(a)-(c) remains non-zero below $h_c$. For a field applied away from the high-symmetry directions, an infinitesimal field immediately triggers a $[\Delta E(\mathbf{k})]_{\mathrm{min}} \neq 0$ (dot-dashed line). Importantly, while the results for the AM-F transition presented
in Figs. \ref{Fig_LineFermi} and \ref{Fig_PhasDiag} refer explicitly to the case of a $B_{1g}^{-}$ AM order parameter
in a tetragonal lattice, they apply to all ordered states of Table
\ref{tab:classification} whose order parameters transform as 1D irreps. 

We note that in certain AM states, besides a magnetic field, strain can also be used to drive the topological AM-F transition even when there is no change in the magnetic ordering. This is because many AM states are piezomagnetic \cite{Aoyama2023,Steward2023}, i.e. strain $\varepsilon_{ij}$ induces a magnetic field component $h_k$ according to $\varepsilon_{ij} = \Gamma_{ijk} h_k$, with the relevant piezomagnetic tensor elements $\Gamma_{ijk}$ being proportional to the AM order parameter $\Phi$. Therefore, in these cases, it is possible to use the appropriate strain components to generate a magnetic field and, consequently, induce the topological AM-F transition. In the specific case of a $B_{1g}^-$ AM order parameter in the tetragonal lattice, in-plane shear strain $\varepsilon_{xy}$ generates the out-of-plane magnetic field $h_z = \gamma \Phi \varepsilon_{xy}$, where $\gamma$ is a coupling constant. The topological transition then takes place when $h_z = h^{(z)}_c$, which corresponds to a critical strain value $\varepsilon_{xy,c} = \frac{\lambda \eta}{\gamma g_{s}\mu_{B}}$ that is independent on $\Phi$.

\section{Conclusions}\label{sec:concl}

In this paper we demonstrated the fundamental role that
the crystalline environment plays on the magnetic and electronic properties of altermagnets.
While in an isotropic altermagnet the magnitude of the magnetization
develops an angular dependence, $\mathbf{M}=d\left(\hat{\mathbf{r}}\right)\left(M_{0,x},\,M_{0,y},\,M_{0,z}\right)$,
in a crystalline environment each component
of the magnetization acquires a different angular dependence, $\mathbf{M}=M_{0}\left(d_{x}\left(\hat{\mathbf{r}}\right),\,d_{y}\left(\hat{\mathbf{r}}\right),\,d_{z}\left(\hat{\mathbf{r}}\right)\right)$.
The resulting group theory classification of the AM order parameters
also suggests new routes to search for altermagnetic states, such
as the $Pnma$ perovskites that realize the $G_{a}C_{b}A_{c}$ magnetic
state. The angular modulation of multiple components of the magnetization in altermagnets leaves clear fingerprints on their electronic
structures, in the form of symmetry-protected nodal lines along crystallographic
mirror planes and Fermi-surface pinch points that behave as type-II Weyl nodes. As a result,
for magnetic fields applied along certain high-symmetry directions
of the crystal, the transition from a nodal to a nodeless Zeeman splitting
of the band structure, which we identify as an AM-F transition, not
only requires a critical magnetic field but is also topological. Beyond
the non-interacting band structure, these symmetry-protected nodal lines
should also impact other electronic properties of altermagnets.  This includes
interaction-driven instabilities towards new electronic states,
such as superconductivity, whose interplay with altermagnets has been recently proposed to be a fertile ground for new phenomena \citep{Mazin2022notes,Ouassou2023,Neupert2023,Sun2023,Papaj2023,Zhu2023,Cano2023}.
\begin{acknowledgments}
We thank J. Sinova and L. \v{S}mejkal for fruitful discussions. RMF was supported by the Air Force Office of Scientific Research under
Award No. FA9550-21-1-0423. TB was supported by the NSF
CAREER grant DMR-2046020. RGP was supported by a grant from the Simons Foundation (Grant Number 1023171) and by the  National Council for Scientific and Technological Development - CNPq.

\end{acknowledgments}

\appendix

\section{\uppercase{Relationship between altermagnetism and multipolar magnetism}} \label{sec_Multipolar}

Here we show that the order parameter of an isotropic AM state is
equivalent to a magnetic multipole moment. Following the main text,
we parametrize the magnetization in the isotropic AM state as:
\be
\mathbf{M}\left(\mathbf{r}\right)=\mathbf{M}_{0}\,Y_{l_{0}m_{0}}\left(\hat{\mathbf{r}}\right),\label{eq:M}
\ee
where $\mathbf{M}_{0}$ is a vector and $Y_{l_{0}m_{0}}\left(\hat{\mathbf{r}}\right)$
is one of the spherical harmonics. We want to compute the magnetic
multipole moments $\mu_{lm}$ of this magnetization function. The
definition is \cite{Hayami2018}:
\be
\mu_{lm}\equiv\sqrt{\frac{4\pi}{2l+1}}\int\mathbf{M}\left(\mathbf{r}\right)\cdot\boldsymbol{\nabla}\left(r^{l}\,Y_{lm}^{*}\left(\hat{\mathbf{r}}\right)\right)\,d^{3}r.
\ee

We assume that the magnetization is contained within a sphere of radius
$R$. For simplicity, we choose $\mathbf{M}_{0}$ to be parallel
to the $z$-axis, i.e. $\mathbf{M}_{0}=M_{0}\hat{\mathbf{z}}$.
Then:
\be
\mu_{lm}\equiv M_{0}\sqrt{\frac{4\pi}{2l+1}}\int Y_{l_{0}m_{0}}\left(\hat{\mathbf{r}}\right)\partial_{z}\left(r^{l}\,Y_{lm}^{*}\left(\hat{\mathbf{r}}\right)\right)\,r^{2}d\Omega.\label{eq:1}
\ee
We have
\begin{align}
\partial_{z}\left(r^{l}\,Y_{lm}^{*}\right) & =\left(\partial_{z}r^{l}\right)Y_{lm}^{*}+r^{l}\left(\partial_{z}Y_{lm}^{*}\right)\nonumber \\
 & =r^{l-1}\left[l\,\cos\theta\,Y_{lm}^{*}-\sin\theta\,\partial_{\theta}Y_{lm}^{*}\right],\label{eq:2}
\end{align}
where we used $\partial_{z}=-\frac{\sin\theta}{r}\,\partial_{\theta}$.
To proceed, it is convenient to write down the formal definition of
$Y_{lm}$ in terms of the associated Legendre polynomials $P_{lm}$:
\be
Y_{lm}^{*}=(-1)^{m}\sqrt{\frac{\left(2l+1\right)\left(l-m\right)!}{4\pi\left(l+m\right)!}}P_{lm}\left(\cos\theta\right)\mathrm{e}^{-im\varphi}.
\ee
To simplify the first term in Eq. (\ref{eq:2}), we use the identity
(hereafter we use a comma to separate the subscripts in order to avoid
confusion):
\be
x\,P_{l,m}\left(x\right)=\frac{\left(l-m+1\right)}{\left(2l+1\right)}P_{l+1,m}\left(x\right)+\frac{\left(l+m\right)}{\left(2l+1\right)}P_{l-1,m}\left(x\right).
\ee
From it, we obtain:
\begin{align}
\cos\theta\,Y_{l,m}^{*} & = \sqrt{\frac{\left(l-m+1\right)\left(l+m+1\right)}{\left(2l+1\right)\left(2l+3\right)}}Y_{l+1,m}^{*} \nonumber\\
& + \sqrt{\frac{\left(l-m\right)\left(l+m\right)}{\left(2l+1\right)\left(2l-1\right)}}Y_{l-1,m}^{*}.\label{eq:2a}
\end{align}

To simplify the second term in Eq. (\ref{eq:2}), we first rewrite
it as
\be
-\sin\theta\,\partial_{\theta}Y_{lm}^{*}=\sin^{2}\theta\,\partial_{(\cos\theta)}Y_{lm}^{*}\underset{x\rightarrow\cos\theta}{=}\left(1-x^{2}\right)\partial_{x}Y_{lm}^{*}.
\ee
We now use the identity:
\begin{align}
\left(1-x^{2}\right)\partial_{x}P_{l,m}\left(x\right) & = \frac{\left(l+1\right)\left(l+m\right)}{\left(2l+1\right)}P_{l-1,m}\left(x\right) \nonumber\\ 
& - \frac{l\left(l-m+1\right)}{\left(2l+1\right)}P_{l+1,m}\left(x\right),
\end{align}
and obtain
\begin{align}
-\sin\theta\,\partial_{\theta}Y_{l,m}^{*} & = \left(l+1\right)\sqrt{\frac{\left(l-m\right)\left(l+m\right)}{\left(2l+1\right)\left(2l-1\right)}}Y_{l-1,m}^{*} \nonumber\\
& - l\sqrt{\frac{\left(l-m+1\right)\left(l+m+1\right)}{\left(2l+1\right)\left(2l+3\right)}}Y_{l+1,m}^{*}.\label{eq:2b}
\end{align}

Substituting Eqs. (\ref{eq:2a}) and (\ref{eq:2b}) in (\ref{eq:2})
gives
\be
\partial_{z}\left(r^{l}\,Y_{l,m}^{*}\right)=r^{l-1}\sqrt{\frac{\left(2l+1\right)\left(l-m\right)\left(l+m\right)}{\left(2l-1\right)}}Y_{l-1,m}^{*}.
\ee
Substituting back in Eq. (\ref{eq:1}) yields
\begin{align}
\mu_{lm} & = M_{0}\sqrt{\frac{4\pi\left(l-m\right)\left(l+m\right)}{\left(2l-1\right)}} \left[\int_{0}^{R}r^{l+1}dr\right] \nonumber\\
& \times \left[\int Y_{l_{0},m_{0}}\left(\hat{\mathbf{r}}\right)Y_{l-1,m}^{*}\left(\hat{\mathbf{r}}\right)d\Omega\right].
\end{align}
The final result is:
\be
\mu_{lm}=\frac{3R^{l_{0}}\mathcal{M}}{l_{0}+3}\sqrt{\frac{\left(l_{0}-m_{0}+1\right)\left(l_{0}+1+m_{0}\right)}{4\pi\left(2l_{0}+1\right)}}\,\delta_{l,l_{0}+1}\delta_{m,m_{0}},
\ee
where, in the last step, we used the fact that $V=\frac{4}{3}\pi R^{3}$
is the volume and defined the total magnetic moment $\mathcal{M}\equiv M_{0}V$.
For a system with uniform magnetization, $l_{0}=0$ and $m_{0}=0$,
such that $M\left(\mathbf{r}\right)=\mathbf{M}_{0}/\sqrt{4\pi}$,
we obtain
\be
\left(\mu_{lm}\right)_{\mathrm{dipole}}=\frac{\mathcal{M}}{\sqrt{4\pi}}\,\delta_{l,1}\delta_{m,0}=\left[\int M\left(\mathbf{r}\right)d^{3}r\right]\delta_{l,1}\delta_{m,0},
\ee
which indeed corresponds to a magnetic dipole moment, since $l=1$.
Therefore, we conclude that the parametrization of the magnetization in Eq. (\ref{eq:M})
corresponds to a multipole magnetic moment of rank $l=l_{0}+1$.

\section{\uppercase{Altermagnetic Landau free-energy expansion}} \label{sec_Landau}

Here we derive the Landau free-energy expansions of the altermagnetic
states described in Table \ref{tab:classification} of the main text. There are three different
types of AM order parameters depending on the irreps under which they
transform: 1D irreps, in which case the AM order parameter has a single
component $\Phi$; 2D irreps, in which case the AM order parameter
has two components $\bm{\Phi}=\left(\Phi^{1},\Phi^{2}\right)$; and 3D
irreps, in which case $\bm{\Phi}=\left(\Phi^{1},\,\Phi^{2},\,\Phi^{3}\right)$. 

The single-component AM order parameter $\Phi$ corresponds to most
of the cases shown in Table \ref{tab:classification} of the main text. In this situation,
using standard methods \cite{Hatch2003INVARIANTS,StokesInvariants}, the Landau free-energy is given by
\be
F=\frac{a}{2}\,\Phi^{2}+\frac{u}{4}\,\Phi^{4}\,,
\ee
and the AM order parameter is Ising-like. There are two cases in which
the AM order parameter has two components: when it transforms as the
$E_{2g}^{-}$ irrep of $D_{6h}$ or as the $E_{g}^{-}$ irrep of $O_{h}$.
In both cases, the Landau free-energy expansion has the same form.
Parametrizing $\bm{\Phi}=\Phi\left(\cos\alpha,\,\sin\alpha\right)$, we
find
\be
F=\frac{a}{2}\,\Phi^{2}+\frac{u}{4}\,\Phi^{4}+\frac{w}{6}\,\Phi^{6}+\frac{\gamma}{6}\,\Phi^{6}\cos6\alpha.
\ee

This is the same Landau free-energy expansion of the six-state clock
model. Minimization with respect to $\alpha$ gives $\alpha_{0}=2n\pi/6$
for $\gamma<0$ or $\alpha_{0}=\left(2n+1\right)\pi/6$ for $\gamma>0$
with $n=0,1,\ldots5$. Alternatively, we can express it as $\alpha_{0}=p\pi/6$,
with $p\in\mathbb{Z}$ even for $\gamma<0$ and odd for $\gamma>0$.

In Table \ref{tab:classification} of the main text, there is only one case in which the AM
order parameter has three components, which correponds to the situation
in which it transforms as the $T_{2g}^{-}$ irrep of $O_{h}$. In
this case, we have $\bm{\Phi}=\left(\Phi^{1},\,\Phi^{2},\,\Phi^{3}\right)$
and the Landau free-energy expansion:

\be
F=\frac{a}{2}\,\Phi^{2}+\frac{u}{4}\,\Phi^{4}+\frac{\gamma}{4}\left(\Phi_{1}^{2}\Phi_{2}^{2}+\Phi_{1}^{2}\Phi_{3}^{2}+\Phi_{2}^{2}\Phi_{3}^{2}\right),
\ee
where $\Phi^{2}\equiv\Phi_{1}^{2}+\Phi_{2}^{2}+\Phi_{3}^{2}$. The
ground state $\bm{\Phi}_{0}$ is given by the sixfold-degenerate manifold $\pm\left(1,\,0,\,0\right)$,
$\pm\left(0,\,1,\,0\right)$, and $\pm\left(0,\,0,\,1\right)$ for
$\gamma>0$, and by the eight-fold degenerate manifold $\pm\left(1,\,1,\,1\right)$, $\pm\left(1,-1,-1\right)$,
$\pm\left(-1,\,1,-1\right)$, and $\pm\left(-1,-1,\,1\right)$, for
$\gamma<0$.

\section{\uppercase{First Principles Methods}}\label{sec_Abinitio_A}

First principles calculations were performed using the Vienna ab Initio Simulation Package (VASP) \cite{kresse1993ab,kresse1996efficiency,kresse1996efficient}. The values reported are calculated using the PBE exchange-correlation functional, and tested that the use of LDA gives comparable results.\cite{PBE} A $16\times 16\times 24$ k-grid for MnF$_2$, and a $8\times 4\times 8$ k-grid for CaMnO$_3$ were used for the reciprocal space integrations. The cutoff energy for plane waves was chosen to be 520 eV, and spin-orbit coupling was taken into account in all calculations. No DFT+U scheme was employed since both compounds prove to be insulating due to the crystal field splitting in the octahedrally coordinated Mn ion. A minimum of 300 electronic steps were considered to be a necessary condition for convergence, since the tiltings of magnetic moments may involve small energy scales.

\section{\uppercase{Symmetry Analysis of Altermagnetic} $Pnma$ \uppercase{Perovskites}}\label{sec_Abinitio_B}

In the standard settings, the $Pnma$ perovskites with the $a^-a^-c+$ rotation pattern has the a and c orthorhombic axes along [$\bar{1}$10] and [110] axes of the pseudocubic unit cell of the perovskite, and the orthorhombic $b$ axis is along the [001] pseudoubic direction, as shown in the figure in the main text. (Note that there is also another common setting, where the long-axis of the orthorhombic cell is the c axis, and the space group name is $Pbnm$.) In this setting, the G-AFM order on the B-site leads to an weak-ferromagnetic moment when its magnetic moments are oriented along the b or c axis, and there is no weak-ferromagnetism only when the magnetic moments of the G-AFM order is oriented along the a axis.

The magnetic space group of this configuration is $Pnma$ (\#62.1.502). Note that while this space group breaks the time-reversal symmetry, it allows no macroscopic magnetic dipole moment since its point group ($mmm$) has three orthogonal mirror planes. However, higher order magnetic multipoles, including octupoles, are allowed, and hence this structure is purely altermagnetic.

\begin{table}[t]
\begin{tabular}{|c|c|c|}
\hline
\hline
        &  & $(x,\; 1/4, \; z \; \vert \; 0, \; m_y, \; 0)$  \\
Ca    & 4c & $(x+1/2, \; 1/4, \; -z+1/2 \; \vert \; 0, \; -m_y, \; 0)$  \\
        &  & $(-x, \; 3/4, \; -z \; \vert \; 0, \; m_y, \; 0)$  \\
        &  &   $(-x+1/2, \; 3/4, \; z+1/2 \; \vert \; 0, \; -m_y, \; 0)$  \\
\hline
        &  & $(0, \; 0, \; 1/2 \; \vert \; m_x, \; m_y, \; m_z)$  \\
Mn   & 4b &      $(1/2, \; 1/2, \; 0 \; \vert \; m_x, \; -m_y, \; -m_z)$  \\
        &  &      $(0, \; 1/2, \; 1/2 \; \vert \; -m_x, \; m_y, \; -m_z)$  \\
        &  & $(1/2, \; 0, \; 0 \; \vert \; -m_x, \; -m_y, \; m_z)$  \\
\hline
        &  & $(x, \; 1/4, \; z \; \vert \; 0, \; m_y, \; 0)$  \\
O     & 4c & $(x+1/2, \; 1/4, \; -z+1/2 \; \vert \; 0, \; -m_y, \; 0)$  \\
        &  & $(-x, \; 3/4, \; -z \; \vert \; 0, \; m_y, \; 0)$  \\
        &  &   $(-x+1/2, \; 3/4, \; z+1/2 \; \vert \; 0, \; -m_y, \; 0)$  \\
\hline
        &  & $(x, \; y, \; z \; \vert \; m_x, \; m_y, \; m_z)$  \\
        &  & $(x+1/2, \; -y+1/2, \; -z+1/2 \; \vert \; m_x, \; -m_y, \; -m_z)$  \\
        &  & $(-x, \; y+1/2, \; -z \; \vert \; -m_x, \; m_y, \; -m_z)$  \\
O     & 8d & $(-x+1/2, \; -y, \; z+1/2 \; \vert \; -m_x, \; -m_y, \; m_z)$  \\
        &  & $(-x, \; -y, \; -z \; \vert \; m_x, \; m_y, \; m_z)$  \\
        &  & $(-x+1/2, \; y+1/2, \; z+1/2 \; \vert \; m_x, \; -m_y, \; -m_z)$  \\
        &  & $(x, \; -y+1/2, \; z \; \vert \; -m_x, \; m_y, \; -m_z)$  \\
        &  & $(x+1/2, \; y, \; -z+1/2 \; \vert \; -m_x, \; -m_y, \; m_z)$  \\
\hline
\hline
\end{tabular}
\caption{Magnetic Wyckoff positions in $Pnma$ CaMnO$_3$.}
\label{tab:CaMnO3Wyckoff}
\end{table}

\begin{figure}[b]
\includegraphics[width=0.40\textwidth]{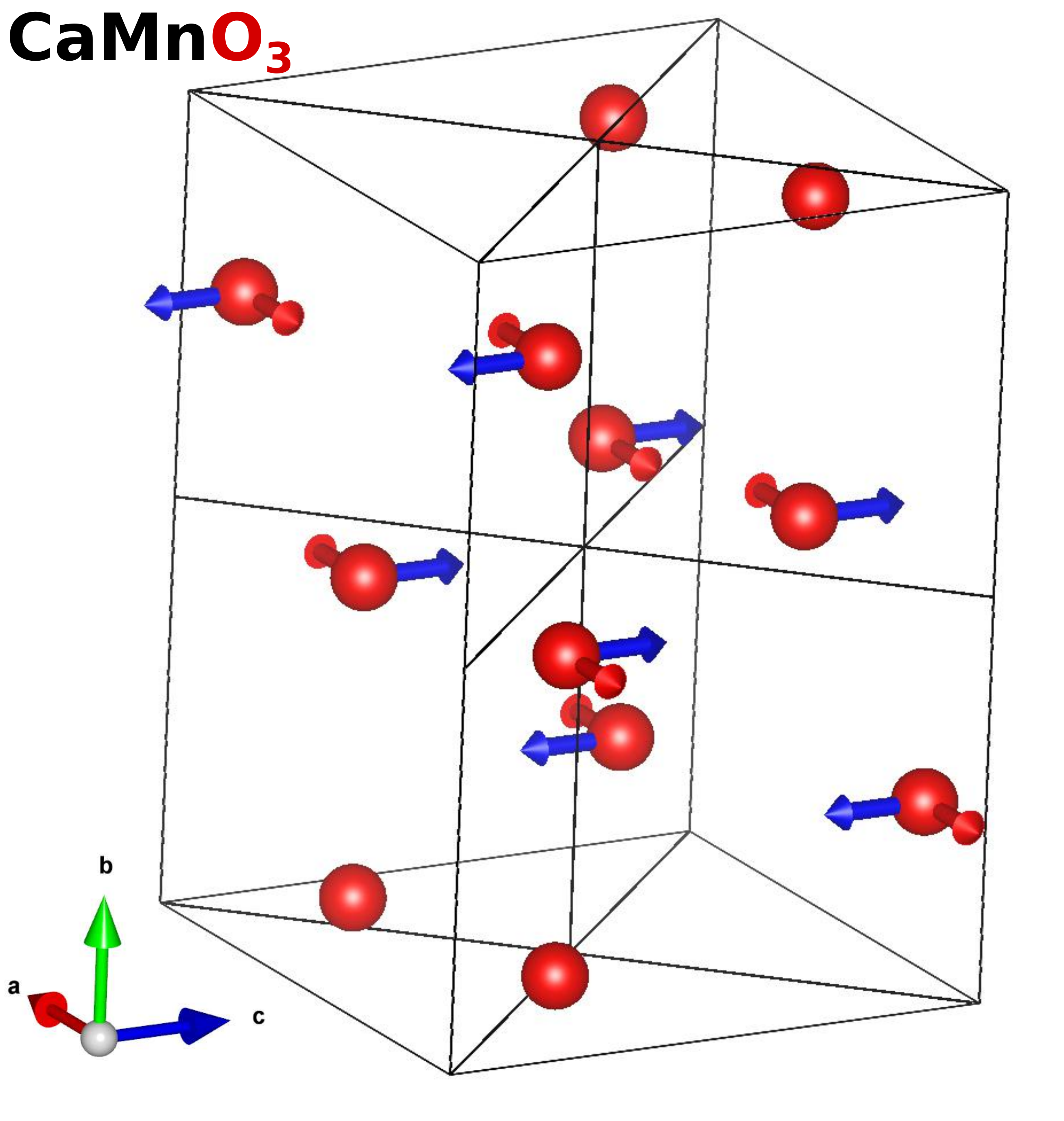}
\label{fig:Pnma}
\caption{Induced antiferromagnetic moments on the O ions on the 8d site in CaMnO$_3$. The magnitudes of magnetic moments along a (green) and c (blue) axes are 0.02 and 0.01 Bohr magnetons respectively. The magnetic moments on the other O ions (Wyckoff letter 4c) and the Ca ions were smaller  than 0.01 Bohr magneton and hence are not shown in the figure.}
\end{figure}

\begin{figure*}[t]
\includegraphics[width=0.70\textwidth]{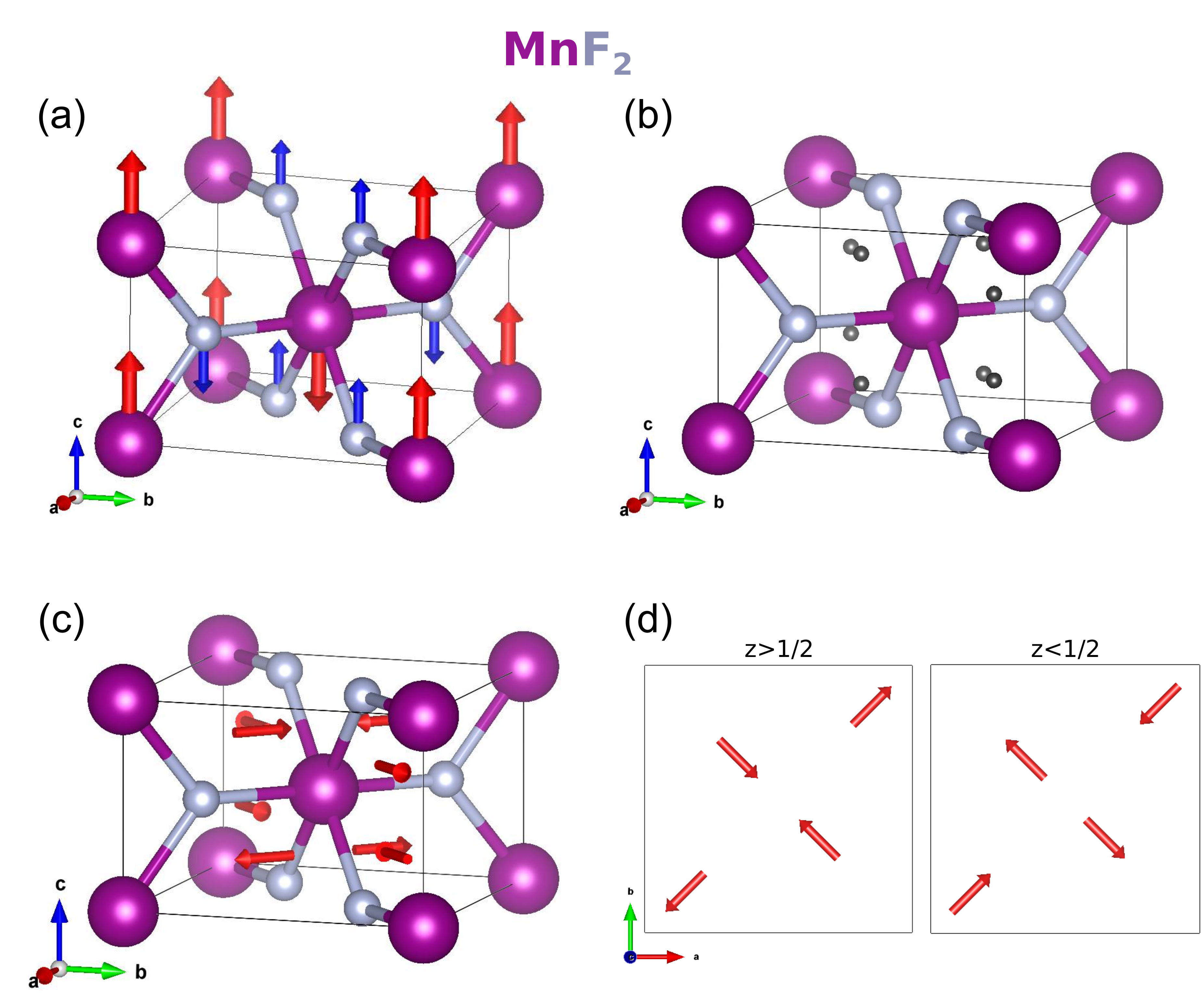}
\caption{(a) MnF$_2$ has the Rutile crystal structure, where the Mn ions (purple spheres) are in the corners and the body center of a simple-tetragonal cell. The AFM order (red arrows) do not reduce the translational symmetry, but induces magnetic moments on the F ions (blue arrows). (b) The 8j Wyckoff sites are shown as grey spheres. (c) In-plane components of the allowed magnetic moments on the 8j Wyckoff sites. (d) Same as panel (c), but from a different angle.}
\label{fig:MnF2}
\end{figure*}

While the presence of higher order magnetic dipoles do not guarantee presence of induced moments on each type of atom, in $Pnma$ perovskites every atom is allowed to have magnetic moments multiple crystallographic directions. This can be best understood using the \textit{magnetic} Wyckoff positions \cite{Gallego2012}. In Table \ref{tab:CaMnO3Wyckoff}, we list the magnetic Wyckoff positions of each atom, along with all the sites, and the magnetic moments on those sites. The Mn atoms are on Wyckoff sites 4b, which allows magnetic moments along all three axes. This means that even though we   initiate DFT calculations with magnetic moments along the $a$ ($x$) axis, the moments on each Mn ion get tilted as the electronic wavefunction is updated. None of the magnetic moments are collinear with each other, but they cancel each other out perfectly due to the multiple mirror and glide planes in the space group, and hence there is no macroscopic dipole moment.

\section{\uppercase{Symmetry Analysis of Altermagnetic M\lowercase{n}F$_2$}}\label{sec_Abinitio_C}

The antiferromagnetic phase of MnF$_2$ [Fig.~\ref{fig:MnF2}(a)], as well as that of RuO$_2$, has the magnetic space group $P4_2'/mnm'$ (\#136.5.1156). Due to the high symmetry of this simple tetragonal group, and the fact that Mn ions are placed on the intersection of multiple mirror planes, their magnetic moments are collinear even in the presence of spin-orbit coupling. This can be seen from the fact that they occupy the Wyckoff site 2a (Table~\ref{table:wyck_mnf2}), which only allows magnetic moments along the c axis. Even though ordered magnetic moments on F ions are induced due to the magnetic order, they are also collinear since the symmetry of the 4f Wyckoff site of F does not allow any noncollinear magnetic moments either. Our first principles calculations show that the magnetic moments on the anions in MnF$_2$ are $\sim 0.03$~$\mu_B$.

\begin{table}[t]
\begin{tabular}{|c|c|c|}
\hline
\hline
        Mn & 2a & $\left(0, \; 0, \; 0 \; \vert \; 0, \; 0, \; m_z\right)$    \\
        && $\left(\frac{1}{2}, \; \frac{1}{2}, \; \frac{1}{2} \; \vert \; 0, \; 0, \; -m_z \right)$\\
\hline
                &       & $\left(x, \; x, \; 0 \; \vert \; 0, \; 0, \; m_z \right)$\\
        F      &  4f  & $\left(-x+\frac{1}{2}, \; x+\frac{1}{2}, \; \frac{1}{2} \; \vert \; 0, \; 0, \; -m_z\right)$\\
                &       & $\left(x+\frac{1}{2}, \; -x+\frac{1}{2}, \; \frac{1}{2} \; \vert \; 0, \; 0, \; -m_z\right)$\\
                &       & $\left(-x, \; -x, \; 0 \; \vert \; 0, \; 0, \; m_z\right)$\\
\hline
                &      & $\left(x, \; x, \; z \; \vert \; m_x, \; m_x, \; m_z\right)$\\
                &       & $\left(-x+\frac{1}{2}, \; x+\frac{1}{2}, \; z+\frac{1}{2} \; \vert \; m_x, \; -m_x, \; -m_z\right)$\\
                &       & $\left(x+\frac{1}{2}, \; -x+\frac{1}{2}, \; z+\frac{1}{2} \; \vert \; -m_x, \; m_x, \; -m_z\right)$\\
         X     &  8j  & $\left(x+\frac{1}{2}, \; -x+\frac{1}{2}, \; -z+\frac{1}{2} \; \vert \; m_x, \; -m_x, \; -m_z\right)$\\
                &       & $\left(-x+\frac{1}{2}, \; x+\frac{1}{2}, \; -z+\frac{1}{2} \; \vert \; -m_x, \; m_x, \; -m_z\right)$\\
                &       & $\left(-x, \; -x, \; z \; \vert \; -m_x, \; -m_x, \; m_z\right)$\\
                &       & $\left(x, \; x, \; -z \; \vert \; -m_x, \; -m_x, \; m_z\right)$\\
                &       & $\left(-x, \; -x, \; -z \; \vert \; m_x, \; m_x, \; m_z\right)$\\
\hline
\hline
\end{tabular}
\caption{Magnetic Wyckoff positions in MnF$_2$ with space group $P4_2'/mnm'$.}
        \label{table:wyck_mnf2}
\end{table}

Even though no noncolinear atomic moments are present in MnF$_2$, this does not mean that the spin density is collinear everywhere in the unit cell. One can use the list of all Wyckoff positions of a magnetic space group, even though they are not occupied by atoms, to predict the direction of the local magnetic moments at a specific point in the unit cell, and whether they lead to different multipoles. In Table~\ref{table:wyck_mnf2}, we list the coordinates and allowed moments of the Wyckoff site 8j, which is shown in Fig.~\ref{fig:MnF2}(b). This unoccupied site allows noncollinear magnetic moments by symmetry, which means that in the AFM phase of MnF$_2$, these points will have a nonzero and noncollinear spin density. The in-plane components of this spin density is shown in Fig.~\ref{fig:MnF2}(c-d). This spin density does not correspond to a single multipole moment, but it is a superposition of many different moments. For example, the in-plane component of the octupole moment $M_{xyz}=\sqrt{15}\left(yz\hat{x} + zx\hat{y} +xy \hat{z}\right)$ is nonzero due to the spin pattern shown in Fig.~\ref{fig:MnF2}.

\section{\uppercase{Symmetry-protected nodal lines}}\label{sec_NodalLines}

Let us analyze the  band degeneracies in the  low-energy   model for AM systems. As explained in the main text, the effective Hamiltonian is given by $\mc H=\sum_{\mb k} \psi^{\dagger}_{\mb k}  H(\mb k)\psi^{\phantom\dagger}_{\mb k}$, where $\psi_{\mb k} = \left( c_{\mb k \uparrow}, \, c_{\mb k \downarrow} \right)^T$ and $  H(\mb k)=  H_0(\mb k)+  H_{\rm int}(\mb k)$ with 
\begin{align}
  H_0(\mb k)& =\varepsilon_{\mb k}  \sigma^0, \\ 
  H_{\rm int}(\mb k)& = -\lambda\mb d_{\rm eff}(\mb k) \cdot\boldsymbol\sigma.
\end{align}
For AM order parameters that transform as 1D irreps of point groups, we have $  \mb d_{\rm eff}(\mb k)=\Phi \, \mb d(\mb k)$ with $\mb d(\mb k)$ given in Table \ref{tab:classification} of the main text.  In the case of multi-dimensional irreps, we define $\mb d_{\rm eff}(\mb k)=\sum_i \Phi^i_0 \, \mb d_i(\mb k)$, where $\Phi^i_0$ is the order parameter configuration that minimizes the corresponding free energy. For the 2D irreps $E_{2g}^-$ in $D_{6h}$ and $E_g^-$ in $O_h$, as shown in Appendix \ref{sec_Landau}, the two-component order parameter is parametrized as in a six-state clock model:
\be
(\Phi^1,\Phi^2)=(\Phi\cos(p\pi/6),\Phi\sin(p\pi/6)),\label{2DPhi}
\ee
where $p\in\{0,1,\dots,11\}$. For the 3D irrep $T_{2g}^-$ in $O_h$, we have two possible configurations, as explained in Appendix \ref{sec_Landau}. The first one is: \be
(\Phi^1,\Phi^2,\Phi^3)=(\Phi\delta_{q1},\Phi\delta_{q2},\Phi\delta_{q3}),\qquad q\in\{1,2,3\},\ee 
whereas the second one is given by
\be\label{NodalT2gOh}
(\Phi^1,\Phi^2,\Phi^3)=\Phi\,(\zeta_1, \,\zeta_2,\zeta_3), 
\ee 
where $\zeta_i = \pm 1$ and $\zeta_1 \zeta_2 \zeta_3 = 1$.

Assuming a ``pure" AM state, which has zero net magnetization, the components of $\mb d_{\rm eff}(\mb k)$ are   homogeneous functions of $\mb k$. The conditions $  d_{\rm eff,\mu}(\mb k)=0$  for each component $\mu\in\{x,y,z\}$ define  sets of planes that contain $\mb k=0$ as well as some high-symmetry directions in reciprocal space.  The band touching (i.e. nodes of the Zeeman splitting)  occurs  when all three components of $\mb d_{\rm eff}(\mb k)$  vanish simultaneously. We denote the nodal lines as follows (here $\alpha\neq\beta\neq\gamma\in\{x,y,z\}$):\bea
\mc L_\alpha&: &\quad  k_\beta=0\quad \forall \beta \neq\alpha, \label{Lalpha}\\
\mc L_{\alpha\beta}&: &\quad  k_\alpha=k_\beta,\quad k_\gamma=0,\\
\mc L_{\overline{\alpha\beta}}&: &\quad  k_\alpha=-k_\beta,\quad k_\gamma=0,\\
\mc L_{6a}&: &\quad  k_y=\sqrt3k_x,\quad k_z=0,\\
\mc L_{\overline{6a}}&: &\quad  k_y=-\sqrt3k_x,\quad k_z=0,\\
\mc L_{6b}&: &\quad  k_y=k_x/\sqrt3, \quad k_z=0, \\
\mc L_{\overline{6b}}&: &\quad  k_y=-k_x/\sqrt3, \quad k_z=0,\\
\mc L_{6c}&: &\quad  k_y=(2+\sqrt3)k_x, \quad k_z=0,\\
\mc L_{\overline{6c}}&: &\quad  k_y=-(2+\sqrt3)k_x, \quad k_z=0,\\
\mc L_{6d}&: &\quad  k_y=(2-\sqrt3)k_x, \quad k_z=0,\\
\mc L_{\overline{6d}}&: &\quad  k_y=-(2-\sqrt3)k_x, \quad k_z=0.\label{L3y}
\eea
In the case of  multi-dimensional irreps, the vector $\mb d_{\rm eff}(\mb k)$ may also display nodal planes, which we denote as\bea
\mc P_{\alpha}&: &\quad   k_\alpha=0.
\eea

\begin{table*}
\caption{\label{Tab_Nodal} Nodal lines and planes in the spectrum of the effective Hamiltonian with AM order parameter transforming as  the irreps of  crystallographic point groups $D_{2h}$, $D_{4h}$, $D_{6h}$, and $O_h$.  In the columns associated with  the nodal lines, as defined in Eqs. (\ref{Lalpha})-(\ref{L3y}), the number 1 (2) indicates that $\mc L$ is a nodal line with Berry phase $\pm \pi$ ($\pm 2\pi$) for the  irrep in the corresponding row.  All nodal lines  belong to at least one  mirror plane of the point group. The check marks in the last three columns indicate that $\mc P_\alpha$ is a nodal plane, in which case the lines contained in that plane are not counted as nodal lines. Note that the nodal lines for the irrep $T_{2g}^-$ with the configuration specified by Eq. \eqref{NodalT2gOh} refer here to the case $\eta = 1$.}
\begin{tabular}{|c|c|c|c|c|c|c|c|c|c|c|c|c|c|c|c|c|c|c||c|c|c|}
\hline  
\hline
Group  & Irrep  & $\mc L_x$ & $\mc L_y$ & $\mc L_z$ & $\mc L_{xy}$&$\mc L_{\overline{xy}}$& $\mc L_{yz}$&$\mc L_{\overline{yz}}$& $\mc L_{xz}$&$\mc L_{\overline{xz}}$& $\mc L_{6a}$&$\mc L_{\overline{6a}}$& $\mc L_{6b}$&$\mc L_{\overline{6b}}$& $\mc L_{6c}$&$\mc L_{\overline{6c}}$& $\mc L_{6d}$&$\mc L_{\overline{6d}}$& $\mc P_{x}$&$\mc P_{y}$& $\mc P_{z}$\tabularnewline 
\hline
$D_{2h}$  & $A_{1g}^-$  & $1$ & $1$ & $1$ &  &  &  &  &  &  &  &  & &  &  &  &  &  &  &  &\tabularnewline 
\hline
$D_{4h}$  & $A_{1g}^-$  & $1$ & $1$ & $1$ & $1$ &$1$  &  &  & &  &  &  & &  &  &  &  &  &  &  &\tabularnewline 
\hline
$D_{4h}$  & $B_{1g}^-$  & $1$ &$1$ & $1$  & &  &  &  & &  &  &  & &  &  &  & &  &  &  &\tabularnewline 
\hline
$D_{4h}$  & $B_{2g}^-$  &  &   &$1$   & $1$ & $1$  &  &  & &  &  &  & &  &  &  & &  &  &  &\tabularnewline 
\hline
$D_{6h}$  & $A_{1g}^-$  & $1$ & $1$  &$1$& &  &  &  & &   &$1$   & $1$ & $1$& $1$ &  &  & &  &  &  & \tabularnewline 
\hline
$D_{6h}$  & $B_{1g}^-$  &  &  &$2$& &  &  &  & &  &  &  & &  &  &  & &  &  &  & \tabularnewline 
\hline
$D_{6h}$  & $B_{2g}^-$  &   & & $2$& &  &  &  & &  &  &  & &  &  &  & &  &  &  & \tabularnewline 
\hline
$D_{6h}$  & $E_{2g}^- \; (p=0,6)$  & 1  & 1 & $1$&   &  &  &  & &  &  &  & &  &  &  & &  &  &  & \tabularnewline 
\hline
$D_{6h}$  & $E_{2g}^- \; (p=1,7)$  &   &  & $1$&    &  &  &  & &  &  &  &  &    & $1$ &  & &  $1$&  &  & \tabularnewline 
\hline
$D_{6h}$  & $E_{2g}^- \; (p=2,8)$  &   &  & $1$&    &  &  &  & &  & $1$  &  & &   $1$&  &  & &    &  &  & \tabularnewline 
\hline
$D_{6h}$  & $E_{2g}^- \; (p=3,9)$  &  &  & $1$&   $1$  &  $1$ &  &  & &  &  &  & &  &  &  & &  &  &  & \tabularnewline 
\hline
$D_{6h}$  & $E_{2g}^- \; (p=4,10)$  &   &  & $1$&    &  &  &  & &  &  &   $1$&  $1$&  & &      &   &  &  &  & \tabularnewline 
\hline
$D_{6h}$  & $E_{2g}^- \; (p=5,11)$  &   &  & $1$&    &  &  &  & &  &  &    &   &  &  &   $1$& $1$ &  &  &  & \tabularnewline 
\hline
$O_{h}$  & $A_{1g}^-$  & $1$  &$1$ &$1$ & $1$&$1$  &$1$  &$1$  &$1$ &$1$  &  &  & &  &  &  &  &  &  &  &\tabularnewline 
\hline
$O_{h}$  & $A_{2g}^-$  &  $1$ &$1$  &$1$ & &  &  &  & &  &  &  & &  &  &  & &  &  &  &\tabularnewline 
\hline
$O_{h}$  & $E_{g}^-\;(p=0,6)$  &   &  & $1$&   &  &  &  & &  &  &  & &   &  &  &  &&  &  &$\checkmark$\tabularnewline 
\hline
$O_{h}$  & $E_{g}^-\;(p=2,8)$  &  $1$ &  &  &    &  &  &  & &  &  &  & &   &  &  &  &&$\checkmark$  &  &\tabularnewline 
\hline
$O_{h}$  & $E_{g}^-\;(p=4,10)$  &    & $1$ &  &    &  &  &  & &  &  &  & &   &  &  &  &&  &$\checkmark$  &\tabularnewline 
\hline
$O_{h}$  & $E_{g}^-\; (p =1,3,5,7,9,11)$  &  $1$ & $1$ & $1$&    &  &  &  & &  &  &   & &  &  &  & &  &  &  & \tabularnewline 
\hline
$O_{h}$  & $T_{2g}^-\;(q=1)$  &  & & $1$  &  $1$  &  $1$ &  &  & &  &  &  & &   &  &  &  &&  &  & \tabularnewline 
\hline
$O_{h}$  & $T_{2g}^-\;(q=2)$  & $1$ &   &   &   &  &  $1$ &  $1$ & &  &  &  & &  & &  &  &  &   &  & \tabularnewline 
\hline
$O_{h}$  & $T_{2g}^-\;(q=3)$  &  & $1$ &   &   &  &  &  & $1$ & $1$  &  &  & &  &   &  &  &  &&   & \tabularnewline
\hline
$O_{h}$  & $T_{2g}^-\; [(1,1,1)]$  &  & &  & $1$ &  & $1$ &  & $1$ &  &  &  & &  &   &  &  &  &&   & \tabularnewline
\hline
$O_{h}$  & $T_{2g}^-\; [(1,-1,-1)]$  &  &  &  & $1$  &  &  & $1$ &  & $1$  &  &  & &  &   &  &  &  &&   & \tabularnewline
\hline
$O_{h}$  & $T_{2g}^-\; [(-1,1,-1)]$  &  &  &   &   & $1$ & $1$ &  &  & $1$  &  &  & &  &   &  &  &  &&   & \tabularnewline
\hline
$O_{h}$  & $T_{2g}^-\; [(-1,-1,1)]$  &  &  &   &   & $1$ &  & $1$ & $1$ &  &  &  & &  &   &  &  &  &&   & \tabularnewline
\hline
\hline
 
\end{tabular}
\end{table*}

In Table \ref{Tab_Nodal} we show the nodal manifolds for the  irreps of the  crystallographic point groups investigated  in this work. For each nodal line, we  compute the Berry phase \be
 \gamma_{\pm }=\oint d\mb k\cdot \mb A_{\pm }(\mb k),
 \ee
 where $\mb A_{\lambda }(\mb k)=i\langle u_{\lambda }(\mb k)|\boldsymbol\nabla_{\mb k}| u_{\lambda }(\mb k)\rangle$ is the Berry connection, calculated from the eigenvectors of the Hamiltonian in the upper ($\lambda=+$) or lower ($\lambda=-$) band, and the integration path encircles the nodal line. On general grounds, we expect topologically trivial  nodal lines ($\gamma_\pm=0$)  to be  unstable against perturbations.

More precisely, we can  analyze the stability of the AM phase using the classification of gapless topological phases \cite{Chiu2015,Schnyder2014,Timm2022}. The band touching in the spectrum of $H(\mb k)$  is governed by  the Fermi ``surface'' (point, line) of $  H_{\rm int}(\mb k)$.   In all cases considered in this work, $  H_{\rm int}(\mb k)$ breaks time-reversal symmetry because $\mb d_{\rm eff}(\mb k)$ is an even function of momentum and time reversal acts in spin space as $\boldsymbol \sigma\mapsto -\boldsymbol \sigma$. However, $  H_{\rm int}(\mb k)$  exhibits a  charge conjugation symmetry \be
\mc C:\quad  C^{-1}  H_{\rm int}(-\mb k)C=-  H_{\rm int}(\mb k),
\ee
with  $C=\mc K\sigma^y$, where $\mc K$ denotes complex conjugation. Since $C^2=-1$, the AM models belong to   class C in the periodic table of topological phases  \cite{Chiu2015}. The nodal lines correspond to  Fermi ``surfaces'' of codimension $2$, whose stability requires mirror symmetries \cite{Schnyder2014}. A mirror symmetry with respect to the plane that contains $\mb k=0$ and is perpendicular to unit vector $\hat{\mb n}$ is defined by the condition\be
R_{\hat{\mb n}} H_{\rm int}(\mc R_{\hat{\mb n}}  \mb k)R_{\hat{\mb n}}=H_{\rm int}(\mb k),\label{mirror}
\ee
where $R_{\hat{\mb n}}=\boldsymbol\sigma\cdot \hat{\mb n}=R_{\hat{\mb n}}^{-1}$ and $\mc R_{\hat{\mb n}} \mb k=-( \mb k\cdot \hat{\mb n})\hat{\mb n}+( \hat{\mb n}\times \mb k)\times  \hat{\mb n}$. If $\mb k$ belongs to the mirror plane, $\mb k\cdot \hat{\mb n}=0$, then $H_{\rm int}(\mb k)$ is invariant under the reflection, and the spin component in the direction of $ \hat{\mb n}$ becomes  a good quantum number.  All  nodal lines listed in Table  \ref{Tab_Nodal} belong to at least one mirror plane.

The perturbation of interest here is the Zeeman term\be
 H_Z=g_s\mu_B\mb h\cdot\boldsymbol\sigma. 
\ee
Note that $H_Z$ respects the charge conjugation symmetry. Mirror symmetry imposes that the magnetic field must be perpendicular to the mirror plane, $\mb h=h\hat{\mb n}$. Importantly, the reflection operator $R_{\hat{\mb n}}$ anticommutes with $C$. In this case, for $h\neq0$ the nodal lines   move away from high-symmetry directions and can be protected by a $2M\mathbb Z$ invariant \cite{Schnyder2014}. The latter can be defined as the difference in  the eigenvalue of $\boldsymbol\sigma\cdot \hat{\mb n}$ for a fixed  band as we vary $\mb k$ across  the  nodal line in the mirror plane. We obtain a  stable AM state   when  the eigenvalue of $\boldsymbol\sigma\cdot \hat{\mb n}$ changes sign with a nontrivial even-parity pattern. This rule generalizes the behavior of the  spin polarization for collinear altermagnets \cite{Smejkal2022_1,Smejkal2022_2}. In addition, we have verified that    symmetry-protected nodal lines are always   associated  with  Berry phase $\pm \pi$ (denoted as 1 in Table  \ref{Tab_Nodal}). In the following we discuss some illustrative examples in detail. We also address the cases of nodal lines with Berry phase $\pm2\pi$ and nodal planes.

\subsection{AM model with  $D_{4h}$ point-group symmetry:    $B_{1g}^-$ irrep}

We begin with the effective Hamiltonian for the AM order parameter that transforms as the $B_{1g}^-$ irrep in the $D_{4h}$ point group: \be
 H(\mb k)=\varepsilon_{\mb k}  \sigma^0 -\lambda\Phi k_yk_z\sigma^x-\lambda\Phi k_xk_z\sigma^y-\lambda\Phi \eta k_xk_y\sigma^z.
 \ee
 Clearly, the spectrum has three nodal lines, $\mc L_x$, $\mc L_y$, and $\mc L_z$. There are three mirror planes, corresponding to $\hat{\mb n}=\hat{\mb x},\hat{\mb y},\hat{\mb z}$. If we apply a magnetic field along an arbitrary direction, all nodal lines are gapped out. To preserve the band touching, the magnetic field must point along the [100], [010], or [001] axes. Consider, for instance, $\mb h=h\hat{\mb z}$.  In this case, we obtain
  \begin{align}
 H(\mb k)+H_Z & = \varepsilon_{\mb k}  \sigma^0 -\lambda\Phi k_yk_z\sigma^x-\lambda\Phi k_xk_z\sigma^y \nonumber\\
 & - (\lambda\Phi \eta k_xk_y-g_s\mu_Bh)\sigma^z.
 \end{align}
 We see that the interacting part of the Hamiltonian no longer vanishes when we set $k_x=k_y=0$, meaning that the degeneracy  along the  $\mc L_z$ line is lifted for $h\neq0$. However, we still obtain a degenerate spectrum for \be
 k_z=0,\qquad \qquad k_xk_y=\frac{g_s\mu_Bh}{\lambda\Phi\eta}.\label{hyperbola}
 \ee
For $h\neq0$, the above equations define two hyperbolic nodal lines contained  in the mirror plane   $k_z=0$. Within this  plane, the Hamiltonian reduces to   \be
 H(k_x,k_y,k_z=0)+H_Z=\varepsilon_{\mb k}  \sigma^0 -(\lambda\Phi \eta k_xk_y-g_s\mu_Bh)\sigma^z.
 \ee
As a result,  the spin polarization  in the $z$ direction becomes a good quantum number. The eigenvalue of $\sigma^z$ in the lower-energy band is given by $\text{sgn}(\lambda\Phi \eta k_xk_y-g_s\mu_Bh )$, changing from $\pm1$  to $\mp1$ as the in-plane momentum crosses the nodal lines.

The two Fermi surfaces of the AM metal touch at four pinch points  where the nodal lines  in Eq. (\ref{hyperbola}) cross the sphere $\varepsilon_{\mb k}=\varepsilon_F$, where $\varepsilon_F$ is the Fermi energy. If we take  the projection of the Fermi surfaces on the mirror plane $k_z=0$,  the   spin polarization changes sign at the pinch points. Note that the vicinity of the pinch points governs the low-energy excitations that involve a spin flip. Assuming  an isotropic  free-electron dispersion $\varepsilon_{\mb k} = k^2/(2m)$, we can calculate the critical value of the magnetic field $h_c^*$ at which  the pinch points annihilate each other. For the field along the $\hat{\mb z}$ direction, we obtain\be
h_c^*=\frac{\lambda \Phi k_F^2}{2g_s\mu_B},
\ee
where $k_F=\sqrt{2m\varepsilon_F}$ is the Fermi momentum.  For $|h|>h_{c}^*$, the nodal lines do not intercept the Fermi sphere, and the system has disconnected Fermi surfaces.

In the limit $|h|\ll h_{c}^*$, we can treat the Zeeman term as a small perturbation and expand the Hamiltonian around the  pinch points with $|\mb k|=k_F$. Consider, for instance, the vicinity of the pinch point on the positive $x$-axis, with $\mb k=(k_F+q_x)\hat{\mb x}+q_y\hat{\mb y}+q_z\hat{\mb z}$, where $|\mb q|\ll k_F$. To first order in $\mb q$, we obtain
\begin{align}
 H(\mb k)+H_Z & \approx \varepsilon_F+  v_Fq_x\sigma^0  -\lambda\Phi k_Fq_z\sigma^y \nonumber\\ 
 & - (\lambda\Phi \eta k_Fq_y-g_s\mu_Bh)\sigma^z,\label{Dirac}
\end{align} 
where $v_F=k_F/m$. The corresponding dispersion reads\be
\varepsilon_{\mb k}\approx \varepsilon_F +v_F q_x\pm \sqrt{(\lambda\Phi k_Fq_z)^2+(\lambda\Phi \eta k_Fq_y-g_s\mu_Bh)^2}. 
\ee
This result is reminiscent of the spectrum of  type-II Weyl semimetals, which   exhibit touching points between electron and hole Fermi pockets \cite{Bernevig2015}. Note, however, that here we have a nodal line, which for $h=0$ runs along the $k_x$ axis.  To leading  order in $h$, the node (pinch point) remains in the $k_z=0$ plane, but  moves in the direction perpendicular to the $k_x$ axis to $q_y=(h/h_c^*)k_F/2$. By taking a cut in momentum space at fixed $q_x=0$, one can see that Eq. (\ref{Dirac}) is equivalent to a Dirac Hamiltonian, which accounts for  the   Berry phase $\pm \pi$ for a path that winds around the nodal line.

\subsection{AM model with $D_{4h}$ point-group symmetry: $A^-_{1g}$ irrep}

\begin{figure}[t]
\centering
\includegraphics[width=0.48\linewidth]{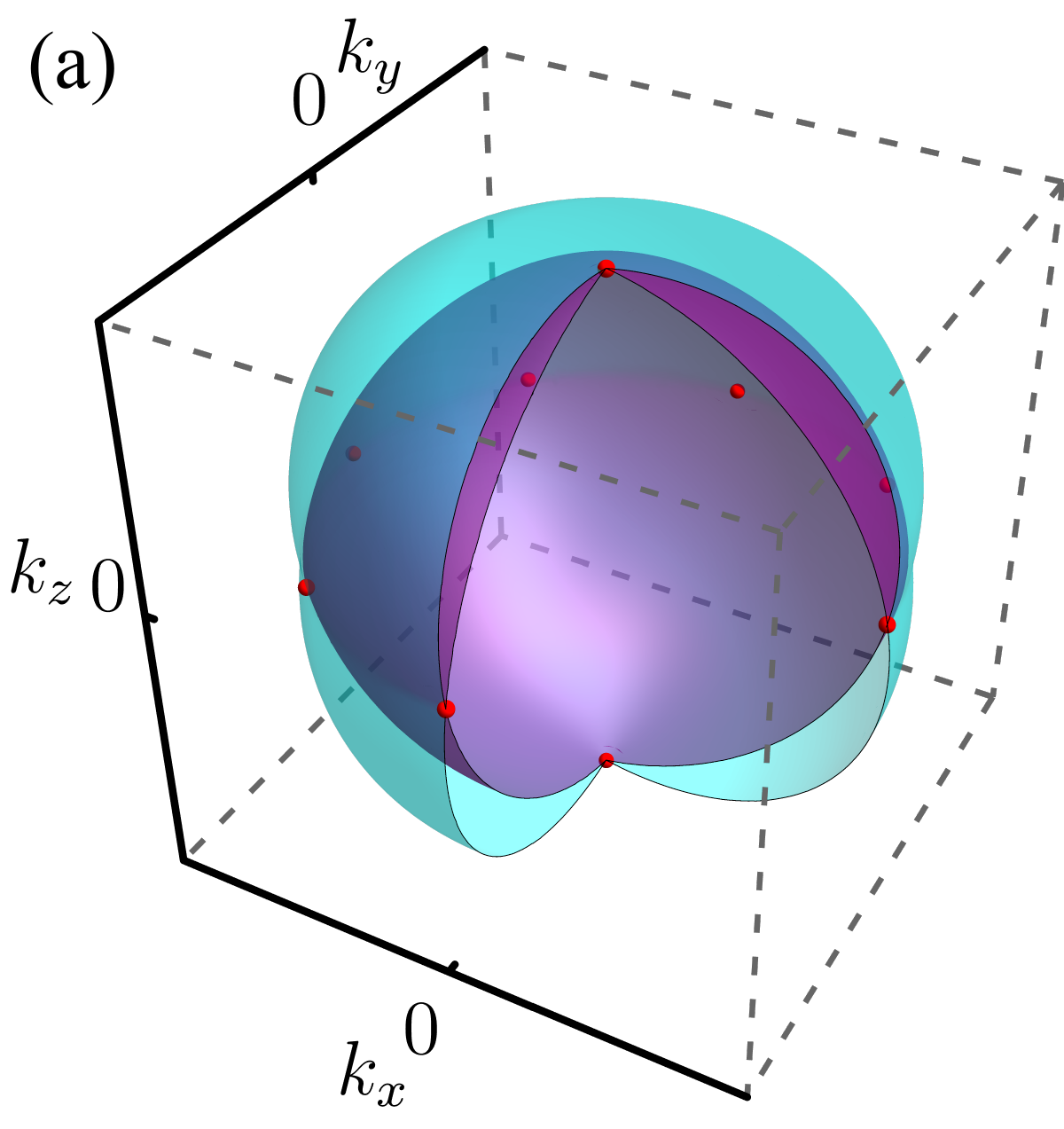} \hfil{} \hfil{} \includegraphics[width=0.48\linewidth]{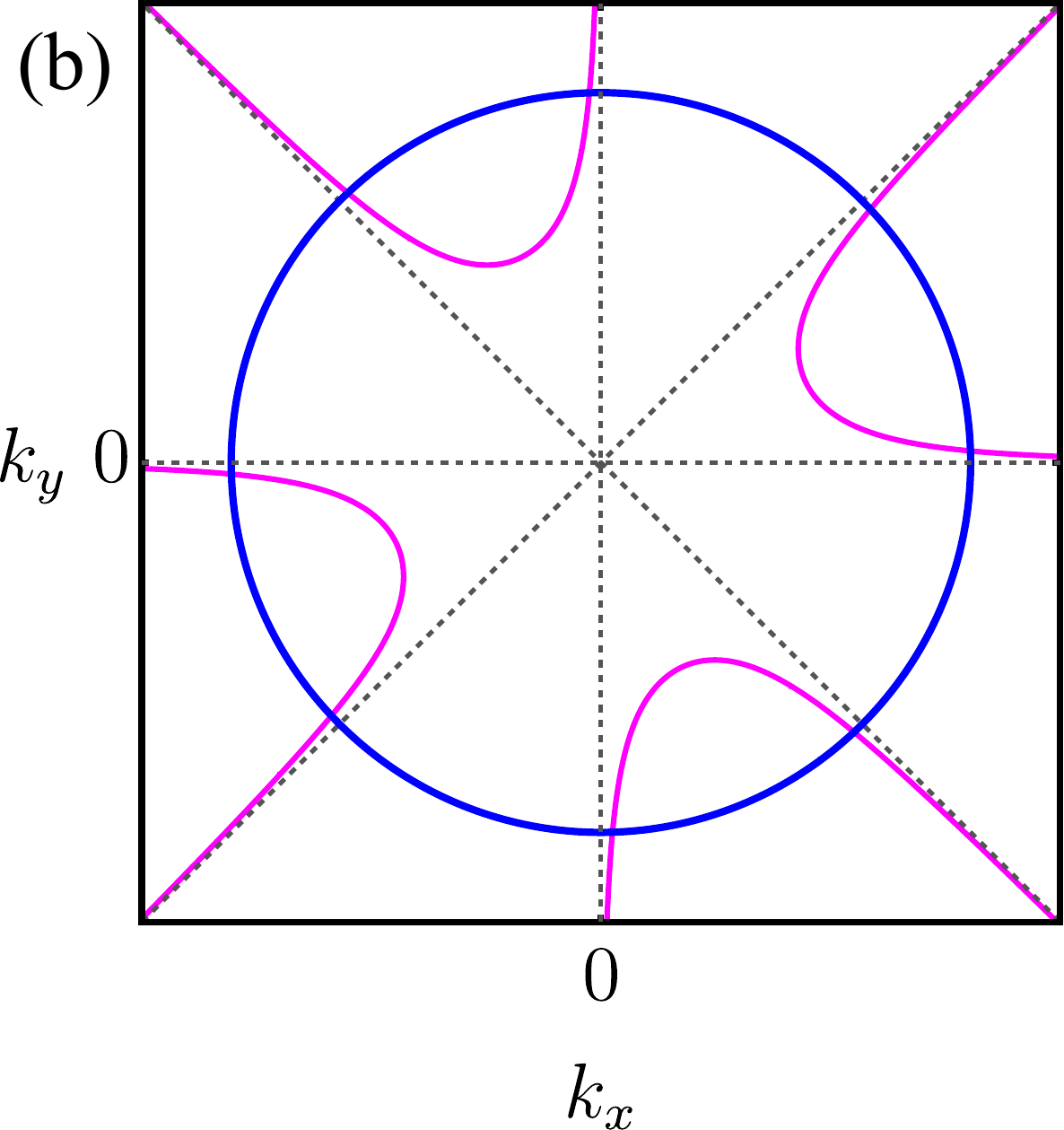}
\caption{AM state whose order parameter transforms as the $A^-_{1g}$ irrep of the $D_{4h}$ group. (a) Fermi surfaces showing 9 out of 10 pinch points (red dots); to aid  visualization, we omit the domain where $k_x>0$ and $k_y<0$.  (b) Nodal lines in the $k_z=0$ plane for $\mb h=0$ (dashed lines) and $\mb h=h\hat{\mb z}$ with $h>0$ (solid magenta lines). The blue circle represents the projection of the noninteracting Fermi surface $|\mb k|=k_F$.  The intersection of the nodal lines with the circle determine the location of the pinch points in the AM phase. }\label{AM_D4h_A1g}
\end{figure}

The effective Hamiltonian in this case reads
\begin{align}
H(\mb k) & =\varepsilon_{\mb k}  \sigma^0 -\lambda\Phi  k_y k_z \sigma^x+\lambda\Phi   k_x k_z\sigma^y \nonumber\\
& -\lambda\Phi \eta k_x k_y(k^2_x - k^2_y)\sigma^z.
\end{align}
As listed in Table \ref{Tab_Nodal},  there are five nodal lines in total,  all of which are associated with Berry phase $\pm \pi$. In the absence of a magnetic field, the Fermi surfaces touch at 10 pinch points; see Fig. \ref{AM_D4h_A1g}(a).  Upon applying a magnetic field in the $z$ direction, the $\mc L_z$ line is gapped out, but the nodal lines in the $k_z=0$  mirror plane persist and are given by the equations\be
k_z=0,\qquad \qquad k_x k_y(k^2_x - k^2_y) =\frac{g_s \mu_B h}{\lambda \Phi\eta}. 
\ee
The nodal lines for $h\geq0$ are depicted in Fig. \ref{AM_D4h_A1g}(b).  Within the $k_z=0$ plane, the Hamiltonian reduces to 
\begin{align}
H( k_x,k_y,k_z=0)+H_Z & =\varepsilon_{\mb k}  \sigma^0  -[\lambda\Phi \eta k_x k_y(k^2_x - k^2_y) \nonumber\\
& - g\mu_sh]\sigma^z.
\end{align}
In the AM phase, the  spin polarization changes sign 8 times as $\mb k$ is varied around one of the  Fermi surfaces in the mirror plane, corresponding to a $g$-wave pattern. The critical magnetic field along the $\hat{\mb z}$ direction is  \be
h_c^*=\frac{\lambda\Phi\eta  k_F^4}{4g_s\mu_B}\qquad \qquad (\mb h\parallel   \hat{\mb z}).
\ee

On the other hand, if we apply the magnetic field along the $\hat{\mb x}$ axis, the AM state has two stable nodal lines in the $k_x=0$ plane, given by \be
k_x=0,\qquad\qquad k_yk_z=\frac{g_s\mu_B h}{\lambda\Phi}. 
\ee
In this case, the critical magnetic field is \be
h_c^*=\frac{\lambda\Phi   k_F^2}{2g_s\mu_B}\qquad \qquad (\mb h\parallel   \hat{\mb x}).
\ee

 \subsection{AM model with $D_{6h}$ point-group symmetry: $B^-_{1g}$ irrep}

According to Table \ref{Tab_Nodal},   the  $B^-_{1g}$ and $B^-_{2g}$ irreps in the $D_{6h}$  group are peculiar  because in these cases there is only one nodal line with Berry phase $\pm 2\pi$. Let us consider the $B^-_{1g}$ irrep. Similar results hold for    $B^-_{2g}$ upon exchanging the roles of $\hat{\mb x}$ and $\hat{\mb y}$ directions. The effective Hamiltonian is 
\begin{align}
H(\mb k) & = \varepsilon_{\mb k}  \sigma^0 -\lambda\Phi  (k^2_x - k^2_y) \sigma^x + 2\lambda\Phi  k_x k_y \sigma^y \nonumber\\
& - \lambda\Phi  \eta k_x k_z(k^2_x - 3 k^2_y)\sigma^z.
\end{align} 
 Let us consider the vicinity of the $\mc L_z$ line. Expanding the Hamiltonian for $\mb k=(q_x,q_y,k_F+q_z)$ to leading order in  $\mb q$, we obtain 
 \be
  H( \mb k)\approx \varepsilon_{F}  \sigma^0 -\lambda\Phi  (q^2_x - q^2_y) \sigma^x + 2\lambda\Phi  q_x q_y \sigma^y , 
  \ee
  where we drop the $\mc O(|\mb q|^3)$ terms   multiplying  $\sigma^z$. This   Hamiltonian describes a  double-Weyl node with $C_6$ symmetry \cite{Bernevig2012}, hence the Berry phase $\pm2\pi$.  Unlike the Dirac Hamiltonian for nodal lines with Berry phase $\pm \pi$, cf. Eq. (\ref{Dirac}), in this case the band splitting  scales  quadratically with momentum in the direction perpendicular to the nodal line.

The $\mc L_z$ line belongs to the mirror plane $k_x=0$.   Adding a Zeeman term with $\mb h=h\hat{\bm x}$, we obtain the Hamiltonian in the mirror plane\be
   H( k_x=0,k_y,k_z)=\varepsilon_{\mb k}  \sigma^0 +(\lambda\Phi  k^2_y+g_s\mu_Bh) \sigma^x .
 \ee
 Note that for $h=0$ the eigenvalue of $\sigma^x$ does not change sign in the mirror plane. As a consequence, the nodal line is not protected by the mirror symmetry. Furthermore, the magnetic field breaks the $C_6$ symmetry of the nodal line, destabilizing the double-Weyl node. The outcome depends on the sign of $h$. For $h>0$, the nodal line is gapped out, but there appear two Weyl points off the mirror plane, located at \be
 k_x=\pm\sqrt{\frac{g_s\mu_B h}{\lambda\Phi}} ,\qquad k_y=k_z=0.
 \ee 
 In this case, finding the band touching at the Fermi level requires fine tuning.  In contrast, for $h<0$ the nodal line splits into two lines   given by\be
 k_x=0,\qquad k_y=\pm \sqrt{\frac{g_s\mu_B|h|}{\lambda\Phi}}.
 \ee
 Note that the two lines remain in the mirror plane, but move off the high-symmetry $k_z$ axis. One can check that, after they split, each line carries a Berry phase $\pm \pi$. Moreover, for $h<0$ the eigenvalue of $\sigma^x$ changes sign across the nodal lines. As a result, we find that for $h<0$ the AM state with four pinch points at the Fermi surface remains stable up to a critical value of the  field \be
 |h_c^*|=\frac{\lambda\Phi k_F^2}{g_s\mu_B}.
 \ee 

  \begin{figure*}[t]
\centering
\includegraphics[width=0.32\linewidth]{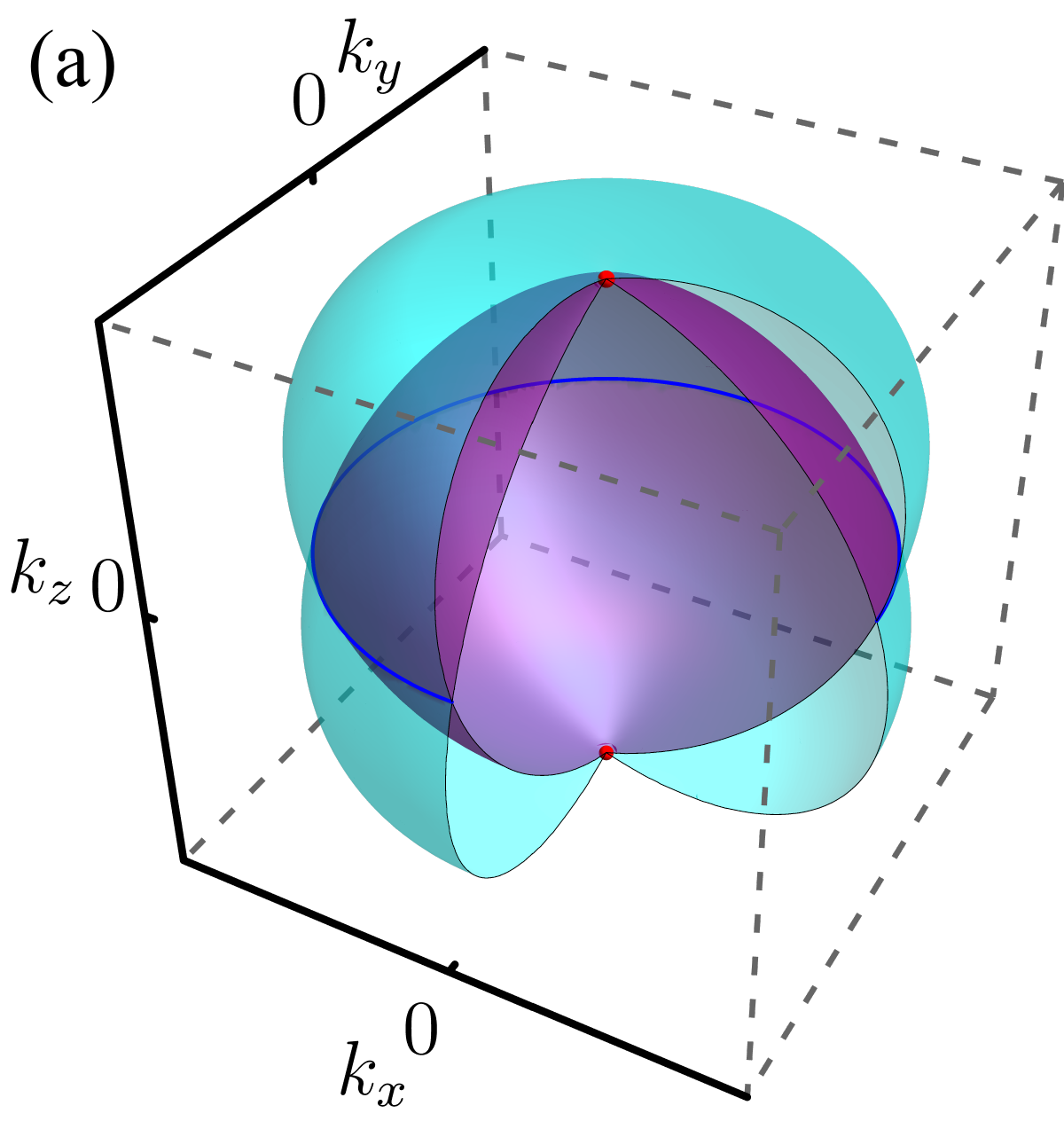} \hfil{} \includegraphics[width=0.32\linewidth]{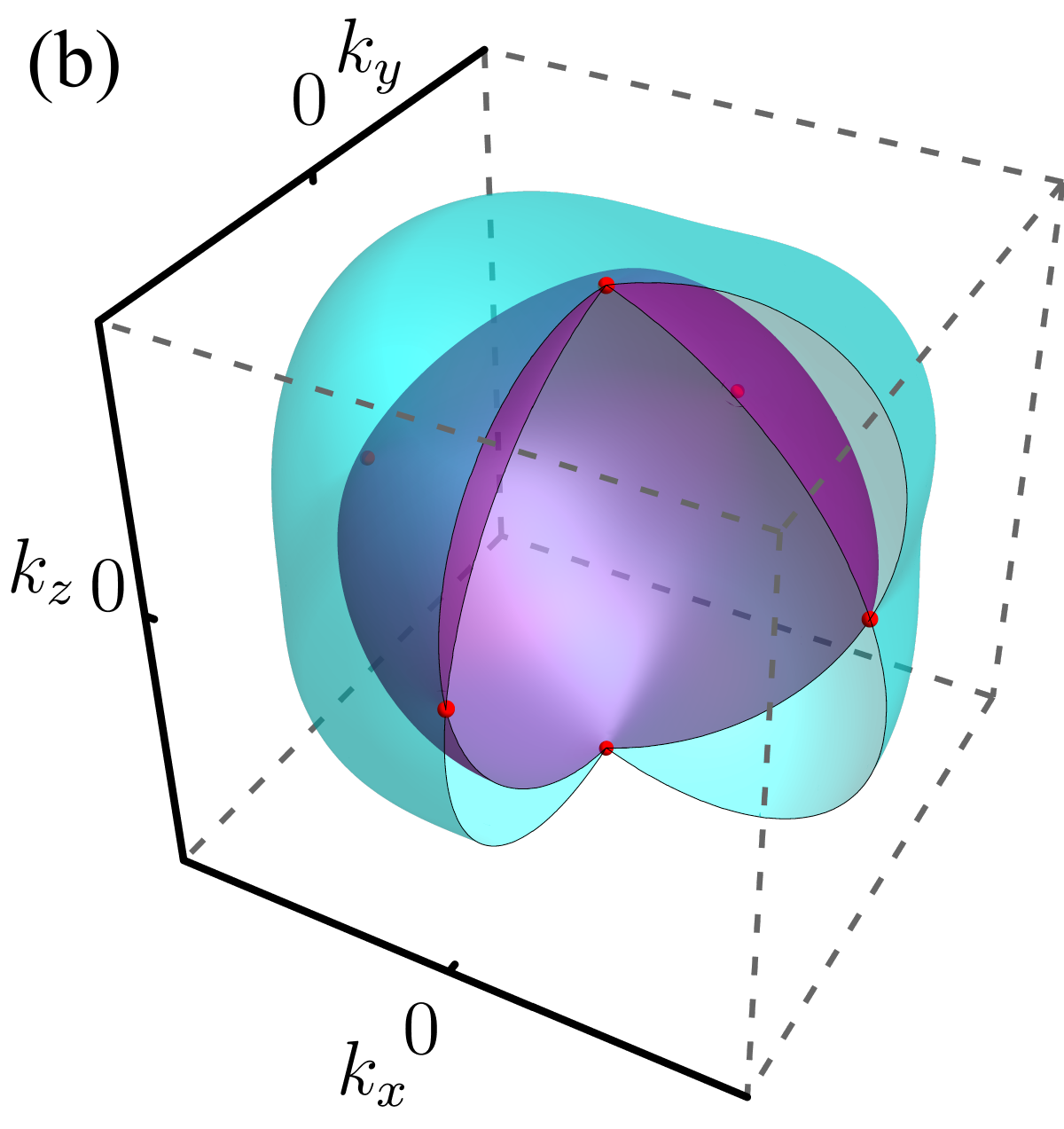} \hfil{} \includegraphics[width=0.32\linewidth]{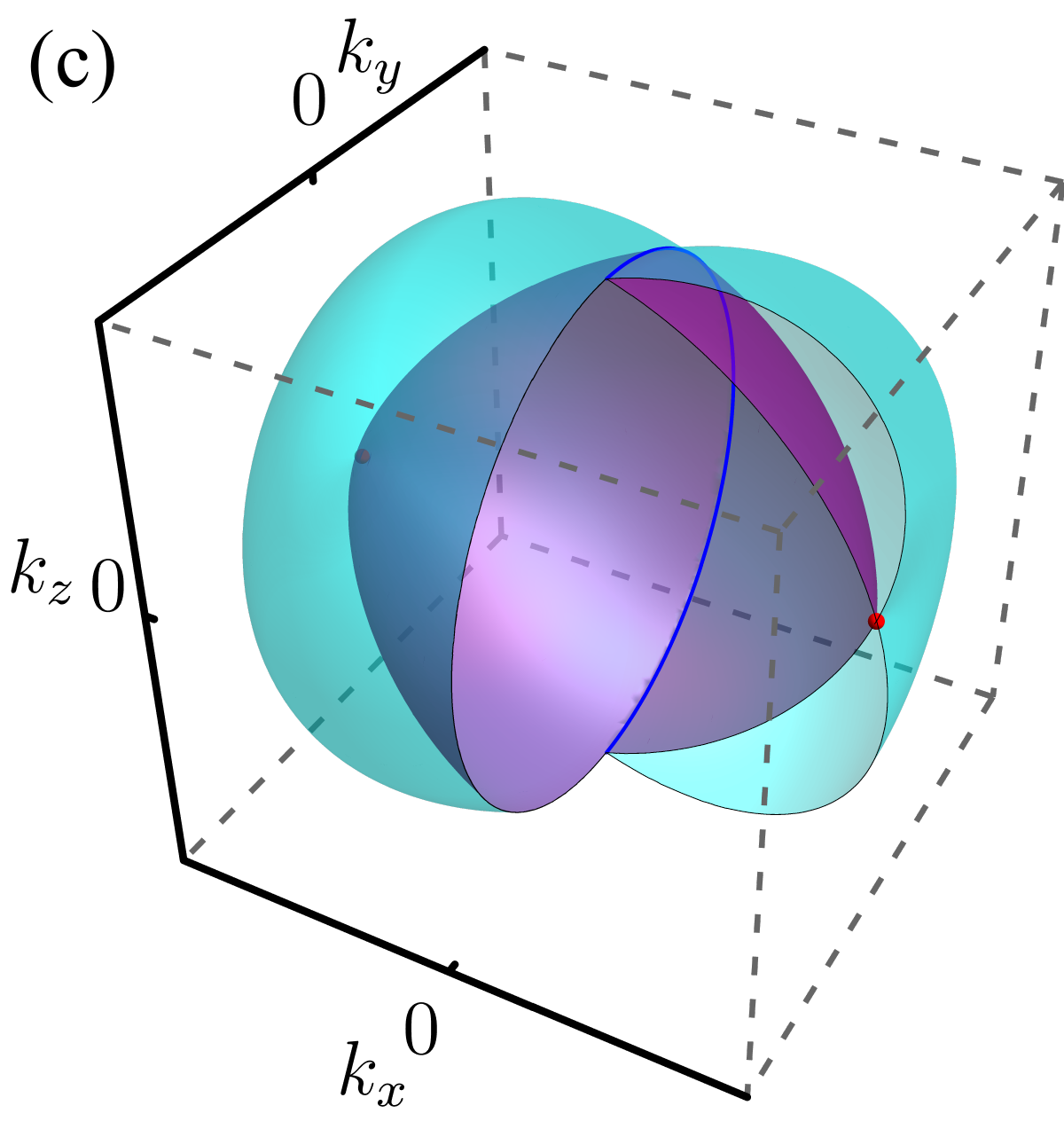}
\caption{Fermi surfaces for the AM state whose order parameter transforms as the $E^-_{g}$ irrep of the $O_{h}$ group; see Eq. (\ref{EgOh}). (a) With the six-state-clock index set to $p=0$, the Fermi surfaces touch at two pinch points associated with  the $\mc L_z$ nodal line and along a line contained in the $\mc P_z$ nodal plane. (b) For $p=1$, the spectrum contains three nodal lines, and the Fermi surfaces touch at 6 pinch points.  (c) The case of $p=2$ is similar to $p=0$, but with Fermi surface touching governed by the $\mc L_x$ nodal line and the $\mc P_x$ nodal plane. }\label{AM_Oh_Eg_FSurface}
\end{figure*}

 \subsection{AM model with $D_{6h}$ point-group symmetry: $E^-_{2g}$ irrep}

We now turn to multi-dimensional irreps, starting with the $E^-_{2g}$ irrep in the $D_{6h}$ point group.  The effective Hamiltonian for the two-component AM order parameter in Eq. (\ref{2DPhi}) is 
\begin{align}
H(\mb k) & = \varepsilon_{\mb k}  \sigma^0 -\lambda\Phi     \left[ \cos \left(\frac{p \pi}{6} \right ) k_y + \sin \left(\frac{p \pi}{6}\right) k_x  \right] k_z\sigma^x \nonumber \\
& -\lambda\Phi      \left[ \cos \left(\frac{p \pi}{6}\right) k_x - \sin \left(\frac{p \pi}{6}\right) k_y \right] k_z \sigma^y \nonumber \\
& -\lambda\Phi\eta \left[ 2\cos\left(\frac{p \pi}{6}\right) k_xk_y    + \sin\left(\frac{p \pi}{6}\right) (k_x^2- k_y^2)  \right] \sigma^z.
 \end{align}
 The location of the nodal lines varies with $p\in\{0,1,\dots,11\}$ as given in Table \ref{Tab_Nodal}. The mirror planes correspond to $\hat{\mb n}=\hat{\mb z}$ and $\hat{\mb n}=\cos\theta_p\,\hat{\mb x}+\sin\theta_p\,\hat{\mb y}$  with $\theta_p=(6-p)\pi/12$. Applying a magnetic field in the $\hat{\mb z}$ direction, we find that the nodal lines in the $k_z=0$ plane are given by 
 \begin{align}
& k_z = 0, \\
& 2\cos\left(\frac{p \pi}{6} \right)k_xk_y+ \sin\left(\frac{p \pi}{6} \right) (k_x^2- k_y^2) = \frac{g_s \mu_B h}{\lambda \Phi \eta}.
 \end{align}
 For all values of $p$, there are two stable nodal lines in the mirror plane. The eigenvalue of $\sigma^z$ changes sign across the nodal lines, and the behavior is similar to the previous examples with Berry phase $\pm \pi$.

\subsection{AM Model with $O_{h}$ point-group symmetry: $E_{g}$ irrep}

As an example of an AM model exhibiting  nodal planes, we consider the  $E_{g}^-$ irrep in the $O_{h}$ point group, with effective Hamiltonian 
\begin{align}
H(\mb k) & = \varepsilon_{\mb k}  \sigma^0 -\lambda\Phi    \left[ \sqrt{3} \cos\left(\frac{p \pi}{6} \right) - \sin\left(\frac{p \pi}{6} \right) \right] k_y k_z \sigma^x \nonumber\\
 & +\lambda\Phi   \left[ \sqrt{3} \cos\left(\frac{p \pi}{6} \right) + \sin\left(\frac{p \pi}{6} \right) \right] k_x k_z \sigma^y \nonumber\\
 & - 2 \lambda\Phi  \sin\left(\frac{p \pi}{6} \right) k_x k_y \sigma^z,\label{EgOh}
\end{align}
where $p\in\{0,1,\dots,11\}$. The mirror planes correspond to $\hat{\mb n}=\hat{\mb x},\hat{\mb y},\hat{\mb z}$. For odd values of $p$, the spectrum contains three nodal lines along the crystallographic axes, and the discussion is similar to the other examples of nodal lines with Berry phase $\pm \pi$. For even values of $p$, the nodal manifolds include a line and the plane perpendicular to it; see Table \ref{Tab_Nodal}. As a result, in the absence of a magnetic field   the Fermi surfaces touch at two pinch points and along a line contained in the nodal plane, as shown in Fig. \ref{AM_Oh_Eg_FSurface}.

To analyze the effects of a magnetic field, let us focus on the Hamiltonian for $p=0$:
 \be
 H(\mb k)=\varepsilon_{\mb k}  \sigma^0 -  \sqrt{3}\lambda\Phi     k_y k_z \sigma^x+  \sqrt{3}\lambda\Phi  k_x k_z \sigma^y.
\ee
Clearly, adding a magnetic field along the $\hat {\mb z}$ direction lifts the band degeneracy, as the $\mc L_z$ nodal line is not protected by the mirror symmetry about the $k_z=0$ plane and nodal planes (codimension 1) in class C are topologically trivial and hence unstable \cite{Schnyder2014}. On the other hand, for a magnetic field along the  $\hat {\mb x}$ direction  we are left with two nodal lines in the $k_x=0$ plane:\be
k_x=0,\qquad \qquad k_y k_z=\frac{g_s\mu_Bh}{\sqrt{3}\lambda\Phi }.
\ee
A similar result holds for a field  along the $\hat {\mb y}$ direction. We conclude that a magnetic field that respects a mirror symmetry and whose direction is contained in the nodal plane turns the nodal plane and the nodal line present at $h=0$ into a pair of nodal lines within the mirror plane.   After the nodal plane is removed for $h\neq0$, the pair of nodal lines in the AM state behaves as in the standard cases.

\bibliographystyle{apsrev4-1_control} 
\bibliography{references_altermagnetism}

\end{document}